\documentclass[aps,pra,twocolumn,amssymb,amsfonts,floatfix,showpacs,superscriptaddress]{revtex4}

\usepackage{graphicx}
\usepackage{dcolumn}
\usepackage{bm}
\usepackage{mathbbol}

\usepackage{amsmath,amssymb,amsthm}
\usepackage{times}
\usepackage{graphicx}
\usepackage{bm}
\usepackage[dvips]{color}

\newcommand{\tr}{\textrm{Tr}}

\begin{document}

\title{Single-particle and many-body analyses of a quasiperiodic integrable system after a quench}

\author{Kai He}
\affiliation{Department of Physics, The Pennsylvania State University, University Park,
Pennsylvania 16802, USA}
\affiliation{Department of Physics, Georgetown University, Washington, DC 20057, USA}
\author{Lea F. Santos}
\affiliation{Department of Physics, Yeshiva University, New York, New York 10016, USA}
\author{Tod M. Wright}
\affiliation{The University of Queensland, School of Mathematics and Physics,
Queensland 4072, Australia}
\author{Marcos Rigol}
\affiliation{Department of Physics, The Pennsylvania State University, University Park,
Pennsylvania 16802, USA}

\begin{abstract}
In general, isolated integrable quantum systems have been found to relax to an apparent equilibrium
state in which the expectation values of few-body observables are described by the generalized Gibbs
ensemble. However, recent work has shown that relaxation to such a generalized statistical ensemble
can be precluded by localization in a quasiperiodic lattice system.  Here we undertake complementary
single-particle and many-body analyses of noninteracting spinless fermions and hard-core bosons
within the Aubry-Andr\'e model to gain insight into this phenomenon. Our investigations span both the
localized and delocalized regimes of the quasiperiodic system, as well as the critical point
separating the two. Considering first the case of spinless fermions, we study the dynamics of the
momentum distribution function and characterize the effects of real-space and momentum-space
localization on the relevant single-particle wave functions and correlation functions.  We show that
although some observables do not relax in the delocalized and localized regimes, the observables that
do relax in these regimes do so in a manner consistent with a recently proposed Gaussian
equilibration scenario, whereas relaxation at the critical point has a more exotic character. We also
construct various statistical ensembles from the many-body eigenstates of the fermionic and bosonic
Hamiltonians and study the effect of localization on their properties.
\end{abstract}

\maketitle

\section{Introduction}
In recent years, there has been a dramatic growth in interest in the physics of nonequilibrium
quantum systems, driven in large part by advances in experimental atomic physics, in particular in
the area of optical lattices~\cite{bloch_dalibard_review_08,Cazalilla2011}.  The precision
time-dependent control and observation of quantum effects afforded by these experiments, together
with the high degree of isolation of the system from the environment, have invigorated the
theoretical study of the time evolution of isolated many-body quantum systems.  The predictions of
previously abstract lines of theoretical inquiry into the mechanisms by which thermal behavior
emerges from purely unitary time evolution, and the role of conservation laws and integrability in
such dynamics, can now be directly compared with empirical evidence acquired in experimental
laboratories.

Several theoretical studies into the quantum origins of thermalization in isolated nonintegrable
systems have found that, away from the edges of the spectrum, few-body observables relax to the
predictions of conventional statistical
ensembles~\cite{Deutsch1991,Srednicki1994,rigol08STATc,rigol09STATa,
rigol09STATb,biroli_kollath_10,neuenhahn_marquardt_12,steinigeweg_herbrych_13}, a phenomenon which
has been connected \cite{Santos2010PRE,Santos2010PREb} to the emergence of quantum chaos
\cite{zele,ZelevinskyRep1996,Flambaum1996,Flambaum1997,BGIC98,I01}.  In addition, it is now well
established that the highly constrained dynamics of integrable systems can in fact give rise to the
relaxation of few-body observables, and that the equilibrium values of these quantities are in many
cases described by the generalized Gibbs ensemble (GGE)
\cite{Rigol2007,rigol_muramatsu_06,cazalilla_06,
calabrese_cardy_07,kollar_eckstein_08,iucci_cazalilla_09,iucci_cazalilla_10,mossel_caux_10,
fioretto_mussardo_10,Cassidy2011,calabrese_essler_11,cazalilla_iucci_12,calabrese_essler_12b,
calabrese_essler_12a,Gramsch2012,ziraldo_silva_12,ziraldo_santoro_13}.

The GGE is constructed by maximizing the many-body entropy \cite{jaynes_57a,jaynes_57b} while
constraining the mean values of all integrals of motion to their expectation values in the initial
state $|\Psi_I\rangle$ of the system. The density matrix in the GGE takes a Gaussian form similar to
that of the grand-canonical ensemble, and can be written as~\cite{Rigol2007}
\begin{equation}\label{eq:dens GGE}
    \hat \rho_{\rm GGE}=\frac{1}{Z_{\rm GGE}} e^{-\sum_s \varLambda_s \hat I_s},
\end{equation}
where  $Z_{\rm GGE}=\tr [e^{-\sum_s \varLambda_s \hat I_s}]$ is the GGE partition function, the $\hat
I_s$ are the conserved integrals of motion, and the $\varLambda_s$ are the corresponding Lagrange
multipliers, which are determined by the constraints $\tr[\hat\rho_{\rm GGE}\,\hat
I_s]=\langle\Psi_I|\hat I_s|\Psi_I\rangle$.

The significance of the fact that the GGE provides an accurate description of observables following
relaxation can be seen by contrasting its predictions with those of the ``diagonal ensemble''
(DE)~\cite{rigol08STATc}: for any initial state $|\Psi_I\rangle$, the time evolution of an observable
$\hat{O}$ under a time-independent (integrable or nonintegrable) Hamiltonian $\hat{H}$ can be written
as
\begin{equation}\label{eq:timeave}
    O(\tau) =\sum_\alpha |C_\alpha|^2 O_{\alpha\alpha} +
\sum_{\alpha\neq\beta} {C^*_\alpha}C_\beta e^{i(E_\alpha-E_\beta)\tau/\hbar} O_{\alpha\beta},
\end{equation}
where $O(\tau)=\langle\Psi (\tau)|\hat{O}|\Psi (\tau) \rangle$, $|\Psi (\tau) \rangle = \sum_{\alpha}
C_{\alpha} e^{-i E_{\alpha} \tau/\hbar} | \psi_\alpha\rangle$,
$\hat{H}|\psi_\alpha\rangle=E_\alpha|\psi_\alpha\rangle$,
$C_\alpha=\langle\psi_\alpha|\Psi_I\rangle$, and $O_{\alpha\beta}=\langle\psi_\alpha|\hat
O|\psi_\beta\rangle$. The infinite-time average of the observable is therefore given by
\begin{align}\label{eq:diagonal}
   \overline{O(\tau)} &= \lim_{\tau'\rightarrow \infty} \frac{1}{\tau'}
\int^{\tau'}_0 d\tau\,  O(\tau) \nonumber\\
&=\sum_\alpha |C_\alpha|^2 O_{\alpha\alpha} \equiv \langle \hat O \rangle_{\rm DE},
\end{align}
which defines the expectation value of $\hat{O}$ in the DE~\cite{rigol08STATc}.
The DE involves as many constraints as the dimension of the many-body Hilbert space (the overlaps
of the initial state with the eigenstates of $\hat{H}$), which grows exponentially with system size.
By contrast, for models that can be mapped to noninteracting Hamiltonians, the GGE involves a number
of constraints that is only polynomially large in the size of the system~\cite{Rigol2007}.
It may, therefore, appear surprising that the predictions of the GGE for expectation values of
observables can agree with those of the DE.

From a many-body perspective, the success of the GGE can be understood as follows~\cite{Cassidy2011}:
the eigenstates of a given integrable Hamiltonian with similar distributions of conserved quantities
have similar expectation values of few-body observables (with the differences vanishing in the
thermodynamic limit). Furthermore, the majority of the states that contribute to the DE have a
distribution of conserved quantities similar (in a coarse-grained sense) to that of the initial
state, and this is also the case for the states that contribute most strongly to the GGE. These facts
imply that differences between the weights in the DE and the GGE are irrelevant and both ensembles
will produce the same results for few-body observables in the thermodynamic limit \cite{Cassidy2011}.
This scenario can be viewed as a generalization of the eigenstate thermalization hypothesis
\cite{Deutsch1991,Srednicki1994,rigol08STATc}, and has been explored in Ref.~\cite{caux_essler_13}
for describing observables after relaxation by means of a single representative state.

Interestingly, it has been recently shown that in an integrable system, in the presence of
localization, the GGE can fail to describe observables after relaxation \cite{Gramsch2012}. This
effect, which parallels the breakdown of eigenstate thermalization in nonintegrable disordered
lattice systems in the presence of localization~\cite{khatami_12}, has been related to the localized
behavior of the underlying system of noninteracting particles to which some integrable models can be
mapped \cite{cazalilla_iucci_12,ziraldo_silva_12,ziraldo_santoro_13}. Here we gain further insights
into this phenomenon by undertaking single-particle and many-body analyses of noninteracting spinless
fermions and hard-core bosons.  We study the dynamics of noninteracting fermions within the
Aubry-Andr\'e model previously studied in Ref.~\cite{Gramsch2012} for hard-core bosons, and show that
although the fermion momentum distribution equilibrates in the localized regime, it fails to
equilibrate in the delocalized one. This should be contrasted with the density profiles, which
exhibit the opposite behavior, equilibrating in the delocalized regime, but not in the localized
regime~\cite{Gramsch2012}. We find that, whenever observables do equilibrate to their GGE values,
they do so in a manner consistent with the Gaussian equilibration scenario of
Ref.~\cite{campos_zanardi_13}. Furthermore, we relate the failure of a given quantity to equilibrate
in a certain regime to the behavior of the single-particle wave functions, as discussed in
Refs.~\cite{cazalilla_iucci_12,ziraldo_silva_12,ziraldo_santoro_13}.

For hard-core bosons, we connect the results of Ref.~\cite{Gramsch2012}, in which the GGE was shown
to fail to describe the momentum distribution function after relaxation in the localized regime, to
the single-particle results for fermions. In addition, for both hard-core bosons and spinless
fermions, we study the density profiles and momentum distribution functions in the many-body
eigenstates of the appropriate Hamiltonians. We focus in particular on the behavior of these
quantities in the many-body eigenstates that contribute to the DE, to the microcanonical ensemble
(ME), and to the microcanonical version of the GGE, i.e., the generalized microcanonical ensemble
(GME)~\cite{Cassidy2011}. We find indications that single-particle real-space localization in the
localized phase and momentum-space localization in the delocalized phase lead to a distinctive
behavior of the many-body eigenstate expectation values of the density and the momentum distribution
of the fermions, respectively. No similar effect is detected in the many-body eigenstate expectation
values of the momentum distribution of the hard-core bosons.

This paper is organized as follows. In Sec.~\ref{Sec:model}, we introduce the models, quench protocols,
and observables studied in later sections. We also review the statistical ensembles utilized
to describe observables after relaxation. The time evolution of spinless fermions following a quench is
studied in Sec.~\ref{Sec:fermions}. Specifically, we examine the relaxation dynamics and time
fluctuations of one-body observables, as well as properties of the single-particle eigenstates that
help us understand the observed out-of-equilibrium behavior. Section~\ref{Sec:hcbs} is devoted to the
study of one-particle observables in the many-body eigenstates of the bosonic and fermionic
Hamiltonians. In Sec.~\ref{Sec:summary}, we summarize our results and present our conclusions.

\section{Models and quenches}
\label{Sec:model}

We consider two models on a one-dimensional lattice with open boundary conditions:
noninteracting spinless fermions (SFs) and hard-core bosons (HCBs).
The HCB model can be mapped onto the model of SFs, by mapping it first onto a spin-1/2 chain via
the Holstein-Primakoff transformation \cite{Holstein1940} and then onto SFs via the
Jordan-Wigner transformation \cite{Jordan1928}.
In both cases, we study the effects of an additional periodic potential,
with a period incommensurate with that of the underlying lattice of the tight-binding model,
which results in the well known Aubry-Andr\'e model~\cite{Aubry1980}. The Hamiltonians for
SFs and HCBs are given by
\begin{eqnarray}
 {\hat H}_f = - t \sum_{j=1}^{L-1} \left( {\hat f}_j^{\dagger} {\hat f}_{j+1} + \text{H.c.} \right)
            + \lambda \sum_{j=1}^L \cos (2 \pi \varsigma j) {\hat n}_j^f ,
\label{HamFermions}
\end{eqnarray}
and
\begin{eqnarray}
 {\hat H}_b = - t \sum_{j=1}^{L-1} \left( {\hat b}_j^{\dagger} {\hat b}_{j+1} + \text{H.c.} \right)
            + \lambda \sum_{j=1}^L \cos (2 \pi \varsigma j) {\hat n}_j^b,
\label{HamBosons}
\end{eqnarray}
respectively, where $L$ is the length of the lattice, ${\hat f}_j$ and ${\hat f}_j^{\dagger}$ (${\hat
b}_j$ and ${\hat b}_j^{\dagger}$) are fermionic (bosonic) annihilation and creation operators on site
$j$, and ${\hat n_j}^f= {\hat f}_j^{\dagger} {\hat f}_j$ (${\hat n}_j^b= {\hat b}_j^{\dagger} {\hat
b}_j$) are SF (HCB) site-occupation number operators.  The prohibition of multiple occupancy of a
single site for HCBs is enforced by the hard-core constraint $\hat b_j^{\dagger 2}=\hat b_j^2=0$.  We
denote the hopping parameter by $t$, and the strength of the incommensurate potential by $\lambda$.
To ensure the incommensurability of the lattice potential, we use an irrational value for
$\varsigma$. We select the inverse golden mean, $\varsigma = (\sqrt{5}-1)/2$, which is considered to
be the most irrational number \cite{Sokoloff1985}. For each given system size, we take the total
number of SFs (HCBs) $N_f$ ($N_b$) to be $N_{f}=N_{b}\equiv N$. In what follows we set $t$ to unity;
i.e., we take $t$ as our unit of energy, and we also set $\hbar=1$ and the Boltzmann constant
$k_B=1$.

The fermionic Hamiltonian (\ref{HamFermions}) is quadratic and therefore trivially solvable: all
many-body eigenstates can be constructed as Slater determinants of the single-particle energy
eigenstates in the incommensurate periodic potentials.  Although the bosonic
Hamiltonian~(\ref{HamBosons}) is also superficially quadratic, the hard-core constraints on the
bosonic creation and annihilation operators encode interactions between the bosons, precluding a
\emph{direct} diagonalization in terms of single-particle states. Nevertheless, it can be solved via
the combined Holstein-Primakoff and Jordan-Wigner transformations, which implies that SF and HCB
systems with Hamiltonians Eqs.~\eqref{HamFermions} and \eqref{HamBosons}, respectively, share the
same (many-body) energy spectrum and consequently have identical thermodynamic properties.  Moreover,
the two models have identical site occupations ${\hat n_j}^f={\hat n}_j^b\equiv{\hat n_j}$.

The properties of HCBs in the Aubry-Andr\'e model have previously been investigated, at both
zero~\cite{rey_satija_06b,he_satija_12} and finite temperature~\cite{nessi_iucci_11}.  The
single-particle Aubry-Andr\'e model exhibits a localization-delocalization transition at the critical
potential strength $\lambda_c = 2$~\cite{Aubry1980}: All single-particle states are extended when
$\lambda < \lambda_c$ and exponentially localized when $\lambda > \lambda_c$.  At the critical point,
the energy spectrum exhibits the fractal structure of a Hofstadter butterfly~\cite{Hofstadter1976}.
As HCBs can be mapped onto noninteracting fermions, they of course inherit this phase transition when
subjected to the incommensurate Aubry-Andr\'e potential. In the localized regime, correlations in the
ground state of the bosonic system decay exponentially with spatial separation, and the system is
said to form a Bose glass \cite{Cazalilla2011}.

Here we study the dynamics and behavior of observables
following sudden quenches of the incommensurate lattice strength $\lambda$.
We choose as our initial states $|\Psi_I\rangle$ the
ground states of initial Hamiltonians ${\hat H}_I$ with parameters $\lambda_I$, and consider the ensuing dynamics
generated by final Hamiltonians ${\hat H}_F$ with parameters $\lambda_F$, corresponding to an instantaneous
change (quench) of the lattice strength from $\lambda_I$ to $\lambda_F$.  We focus on one-body observables
that are accessible in optical lattice experiments: the density profiles ${\hat n}_j$,
and the momentum distribution functions
\begin{eqnarray}
&&\hat{m}^f_k = \frac{1}{L} \sum_{j,j'=1}^L e^{i k (j-j') } {\hat f}_j^{\dagger} {\hat f}^{}_{j'}
\label{momFermions}
\end{eqnarray}
and
\begin{eqnarray}
&&\hat{m}^b_k = \frac{1}{L} \sum_{j,j'=1}^L e^{i k(j-j') }  {\hat b}_j^{\dagger} {\hat b}^{}_{j'} .
\label{momBosons}
\end{eqnarray}
We note that although the site
occupations are equal for SFs and HCBs, the off-diagonal spatial correlations,
and therefore the momentum distributions, of the two systems are distinct.
To calculate $\langle \hat{m}^b_k\rangle$ in a pure state, we follow the approach of
Refs.~\cite{Rigol2004,Rigol2005,He2011}, while for
calculations in the GGE we use the methodology of Ref.~\cite{rigol_05}.

\subsection*{Ensembles of interest}
We characterize the behavior of our quasiperiodic system after relaxation following a quench by
comparing it to three different statistical ensembles: DE, ME, and GME, each of which we briefly
describe here.

{\it Diagonal ensemble.}  The density matrix of the DE is defined by
\begin{equation}
\hat{\rho}_\text{DE}= \lim_{\tau'\rightarrow \infty} \frac{1}{\tau'}
\int^{\tau'}_0 d\tau\, |\Psi (\tau) \rangle\langle\Psi (\tau)|
=\sum_\alpha |C_\alpha|^2 |\psi_\alpha \rangle\langle\psi_\alpha|,
\end{equation}
i.e., it is diagonal in the (many-body) energy representation, and the energy eigenstates are weighted
according to their overlaps with the initial state.  The expectation value of an observable in
this ensemble is given by Eq.~\eqref{eq:diagonal}.  In various computational studies
\cite{rigol09STATa, rigol09STATb,Cassidy2011}, observables after relaxation have been shown to
approach the DE predictions [Eq.~\eqref{eq:diagonal}] as the system size increases.

{\it Microcanonical ensemble.}  The density matrix of the ME can be written
as
\begin{equation}
\hat{\rho}_\text{ME}= \frac{1}{{\cal N}_{E,\delta_{\textrm{ME}}}}
\sum_{\underset{ |E-E_{\alpha}| < \delta_{\textrm{ME}} }{\alpha}}|\psi_\alpha \rangle\langle\psi_\alpha|;
\end{equation}
i.e., all eigenstates in the energy window $[E-\delta_{\textrm{ME}}, E+\delta_{\textrm{ME}} ]$
(of which there are ${\cal N}_{E,\delta_{\textrm{ME}} }$) are given equal weight.
We have checked that expectation values of observables within
our microcanonical calculations are robust against the exact value of $\delta_{\textrm{ME}}$. To ensure
this, we select $\delta_{\textrm{ME}}$ to be much smaller than the full spectrum width, but
sufficiently large to contain many eigenstates (i.e., larger than the average level spacing at the given
$E$). In general, $\delta_\mathrm{ME}=0.05$ for most results reported in this work.

{\it Generalized Gibbs ensemble.} The density matrix for this ensemble,  which is of a similar
Gaussian form to that of the grand-canonical ensemble, was already introduced in Eq.~\eqref{eq:dens
GGE}. We note that a recipe for constructing the appropriate conserved quantities, allowing for the
extension of the GGE description to more general systems than those considered in this article, has
recently been proposed~\cite{OlshaniiARXIV}.  However, we make here the ``natural'' choice for the
conserved quantities, taking them to be the occupations of the single-particle eigenstates of the
noninteracting SFs \cite{Rigol2007}.

{\it Generalized microcanonical ensemble.} The GME is the microcanonical  version of the GGE
\cite{Cassidy2011}. The only energy eigenstates that contribute to this ensemble are those that have
distributions of the conserved quantities that are similar (in a coarse-grained sense) to that of the
initial state. These eigenstates are all assigned the same weight, as in the usual ME. The
density matrix of the GME therefore has the form
\begin{equation}
\hat{\rho}_\textrm{GME} = \frac{1}{{\cal N}_{\{ I_s \},\delta_{\textrm{GME} } } }
\sum_{\underset{ \delta_{\alpha } < \delta_\textrm{GME}}{\alpha}} |\psi_\alpha \rangle\langle\psi_\alpha|,
\label{Ogme}
\end{equation}
where $\delta_{\alpha}$ measures the distance of the eigenstate $|\psi_\alpha \rangle $ from
the target distribution of conserved quantities determined by the initial state,
and
${\cal N}_{\{ I_s \},\delta_\textrm{GME} } $ is the number  of energy eigenstates within the
GME window $\delta_{\alpha} < \delta_\textrm{GME}$.

In order to construct the GME, one needs to compare the distribution of conserved quantities in each
of the eigenstates of ${\hat H}_F$ with that of the initial state.  Since the conserved quantities
are fermion occupation numbers of single-particle energy eigenstates, their expectation values
$I_{s,\alpha}=\langle\psi_\alpha| \hat{I}_s|\psi_\alpha\rangle$ in the many-body eigenstates are
either 0 or 1. By contrast, the occupations of the conserved quantities in the initial state (which
are equal to those in the DE) can assume any real value between these two values; i.e., $0
\leq\langle\hat{I}_s\rangle_\textrm{DE}=\sum_{\alpha}|C_{\alpha}|^2 I_{s,\alpha}\leq1$. To compare
those distributions, in a coarse-grained way, we proceed as follows \cite{Cassidy2011}.

(i) We sort the conserved quantities so that $\langle \hat{I}_s \rangle_\textrm{DE}$ decreases
monotonically as $s$ increases. In this way we obtain a comparatively smoothly varying discrete
distribution suitable for coarse graining.

(ii) After sorting, we generate a discrete target distribution of conserved quantities from $\langle
\hat{I}_s \rangle_\textrm{DE}$. This is achieved by interpolating $\langle \hat{I}_s \rangle_\textrm{DE}$ to find a
continuous function $I(s)$ (where $s$ can now be any real number in the interval [$0.5,L+0.5$]), and
then computing all values $s_l^{*}$ satisfying $\int_{0.5}^{s_1^*} I(s)\, ds = 0.5$ and
$\int_{s_{l-1}^*}^{s_l^*} I(s) \,ds = 1$ for $l>1$. The set $\{ s_l^{*} \}$, together with the
set of corresponding weights $\{ I(s_l^*) \}$, defines the target distribution.

(iii) We introduce a measure $\delta_{\alpha}$ to quantify how close the distribution of each many-body
eigenstate $|\psi_\alpha\rangle$ is to the target distribution. Those states with $\delta_{\alpha}
<\delta_\textrm{GME}$ are included in the GME. The choice of $\delta_{\alpha}$ is not unique. Following
Ref.~\cite{Cassidy2011}, we choose
\begin{equation}
\delta_{\alpha} = \sqrt{ \frac{1}{N} \sum_{l=1}^{N}  I(s_l^*)   ( s_{l,\alpha}^{}  -  s_l^{*})^2 } ,
\label{deltaA}
\end{equation}
where $s_{l,\alpha}^{}$, with $l=1,  \ldots, N$, enumerate the sorted single-particle states [see
(i)] occupied in eigenstate $|\psi_\alpha\rangle$. We choose the value of $\delta_\textrm{GME}$ to be that
which yields the minimum value of the normalized absolute difference
\begin{equation}\label{eq:madD}
 D= \frac{\sum_{s=1}^{L}|\langle \hat{I}_s\rangle_\textrm{GME}-\langle\hat{I}_s\rangle_\textrm{DE}|}
{\sum_{s=1}^{L}\langle \hat{I}_s\rangle_\textrm{DE}},
\end{equation}
between the expectation values of the conserved quantities in the GME and DE.
As in our microcanonical calculations, we have checked that
our results are robust against small changes in the value of $\delta_\textrm{GME}$.

\section{Single-particle analysis}
\label{Sec:fermions}

In Ref.~\cite{Gramsch2012} it was shown that, following a quench of HCBs to the localized regime of
the Aubry-Andr\'e model, one-body observables that depend on nonlocal correlations, such as $m^b_k$,
do relax to time-independent values (with fluctuations vanishing in the thermodynamic limit), but
these values are not consistent with the predictions of the GGE.  This should be contrasted with the
on-site density, whose time average agrees with the GGE results in all regimes. It was also found
that the dynamics of $n_j$ and $m^b_k$ are qualitatively different in the delocalized and localized
regimes: the momentum distribution $m^b_k$ approaches a time-independent value with increasing system
size regardless of whether the system is in the delocalized or localized regime, while $n_j$ only
exhibits such relaxation in the delocalized regime. All results for the density also apply to SFs, to
which HCBs can be mapped.

Here, we begin by studying the relaxation dynamics of the momentum distribution $m_k^f$ of SFs,
which is in general completely unrelated to the momentum distribution of the corresponding system of
HCBs. In fact, the GGE describes, by construction, the infinite-time averages of all one-body fermionic
observables regardless of whether the single-particle states are localized or delocalized and
independently of the system size. This can be straightforwardly proven by projecting
$\hat{\rho}(\tau)=|\Psi(\tau)\rangle\langle \Psi(\tau)|$ onto the single-particle sector. Considering
that all eigenstates of the many-body Hamiltonian are (antisymmetrized) direct products of the single-particle
states $|s\rangle$ in which $\hat{H}_f$ is diagonal ($\hat{H}_f|s\rangle=e_s|s\rangle$),
the time evolution of the one-particle density matrix can be cast in the form
\begin{equation}\label{eq:singt}
 \hat{\rho}_\text{sp}(\tau)=\sum_{s,s'} c_{ss'} e^{-i(e_s-e_{s'})\tau}|s\rangle\langle s'|.
\end{equation}
In the absence of degeneracies in the  single-particle spectrum (which is the case in the
Aubry-Andr\'e model considered here), the infinite-time average of $\hat{\rho}_\text{sp}(\tau)$
can be written as
\begin{equation}
\overline{\hat{\rho}_\text{sp}(\tau)}=
\lim_{\tau'\rightarrow \infty} \frac{1}{\tau'}
\int^{\tau'}_0 d\tau\, \hat{\rho}_\text{sp}(\tau)
=\sum_s c_{ss} |s\rangle\langle s|,
\end{equation}
which is, by construction, the single-particle density matrix predicted by the GGE, as
$c_{ss}=\sum_{\alpha}|C^{}_{\alpha}|^2 I_{s,\alpha}\equiv\tr [\hat \rho_{\rm GGE}\, \hat I_s]$.
We emphasize, however, that this does not necessarily imply that all such observables
exhibit \emph{relaxation} to their GGE values.  In particular, the results of
Ref.~\cite{Gramsch2012} indicate that the on-site densities in the localized phase exhibit
finite fluctuations about their GGE expectation values even in the thermodynamic limit.

\begin{figure}[!t]
  \centering
  \includegraphics[width=0.475\textwidth]{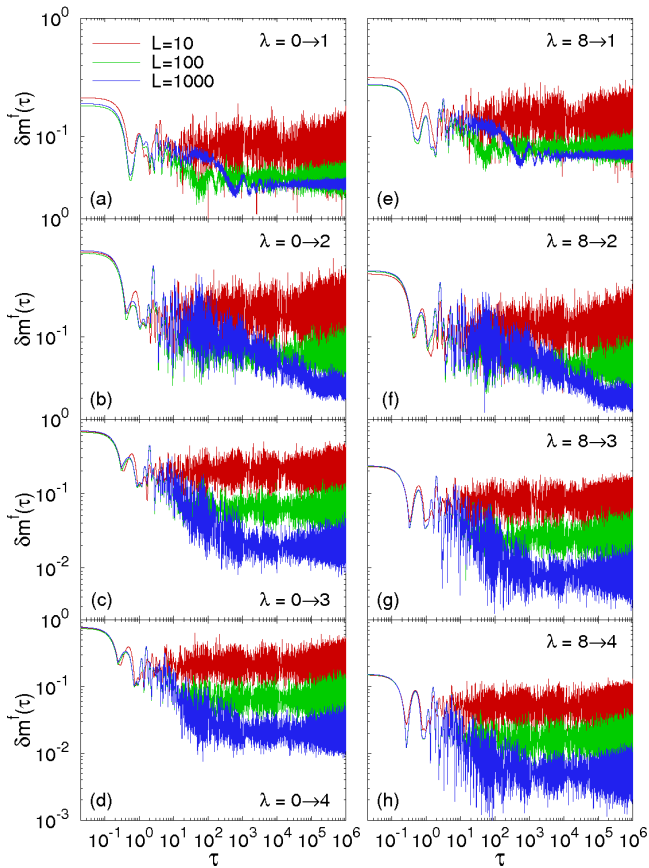}
  \caption{(Color online) Relaxation dynamics of $m^f_k$ in quenches
  $\lambda=\lambda_I \rightarrow \lambda_F $ (as indicated in the panels)
for systems with 10, 100, and 1000 lattice sites (curves from top to bottom in each panel), and $N=L/2$.}
\label{fig:fermdyn}	
\end{figure}

In order to quantify how closely $m_k^f(\tau)=\langle \Psi(\tau)|\hat{m}_k^f|\Psi(\tau)\rangle$
approaches the corresponding GGE prediction, we compute the normalized difference
\begin{equation} \label{eq:deltaO}
  \delta m^f(\tau)=\frac{\sum_{k}|m_k^f(\tau)-\langle \hat{m}_k^f\rangle_\text{GGE}|}
                      {\sum_{k}\langle \hat{m}_k^f\rangle_\text{GGE}},
\end{equation}
between the instantaneous momentum distribution at time $\tau$ and the GGE prediction for this
quantity. Relaxation of $m_k^f$ to the GGE prediction is observed if $\delta m^f (\tau)$ vanishes at
long times, in the thermodynamic limit. In practical numerical calculations, however, $\delta
m^f(\tau)$ will always be finite, because of finite-size effects. The signature of relaxation to the
GGE in our calculations is therefore that $\delta m^f (\tau)$ fluctuates about a finite average value
at long times, and that this average value scales towards zero with increasing system size.

In Fig.~\ref{fig:fermdyn}, we show results for $\delta m^f(\tau)$ in quenches with $\lambda_I=0$;
i.e., a delocalized initial state (left panels), and $\lambda_I=8$; i.e., a localized initial state
(right panels). After the quench, $\lambda_F=1$ [delocalized regime, Figs.~\ref{fig:fermdyn}(a) and
\ref{fig:fermdyn}(e)], $\lambda_F=2$ [critical point, Figs.~\ref{fig:fermdyn}(b) and
\ref{fig:fermdyn}(f)], and $\lambda_F=3$ and 4 [localized regime, Figs.~\ref{fig:fermdyn}(c),
\ref{fig:fermdyn}(g), and \ref{fig:fermdyn}(d), \ref{fig:fermdyn}(h), respectively]. The results
presented correspond to three different system sizes ($L=10,\,100,$ and 1000, curves from top to
bottom in each panel, and $N=L/2$), and to the same quenches and system sizes studied for HCB systems
in Ref.~\cite{Gramsch2012}.

The results obtained for a given final value of the incommensurate potential strength are
qualitatively similar, independently of whether the initial state is delocalized or localized. In
quenches to the localized regime [Figs.~\ref{fig:fermdyn}(c), \ref{fig:fermdyn}(d),
\ref{fig:fermdyn}(g), and \ref{fig:fermdyn}(h)], we observe that $\delta m^f(\tau)$ decays to a
finite value, about which it undergoes fluctuations, and that this value decreases with increasing
system size. In quenches to the delocalized regime
[Figs.~\ref{fig:fermdyn}(a)~and~\ref{fig:fermdyn}(e)], $\delta m^f(\tau)$ similarly undergoes decay
to a finite value about which it fluctuates.  However, in the delocalized case, the value to which
$\delta m^f(\tau)$ decays does not appear to exhibit such a pronounced reduction as the system size
is increased, suggesting that it may not tend towards zero as $L\to \infty$. Following quenches to
the critical point [Figs.~\ref{fig:fermdyn}(b) and \ref{fig:fermdyn}(f)], $\delta m^f(\tau)$ exhibits
behavior similar to that observed in the localized regime, decaying to exhibit fluctuations about a
constant value that decreases with increasing system size. However, in this critical regime the decay
of $\delta m^f(\tau)$ is much slower and, e.g., fluctuation about a constant value in the case
$L=1000$ is only obtained for $\tau\gtrsim10^5$.

\begin{figure}[!t]
	\centering
	\includegraphics[width=0.4\textwidth]{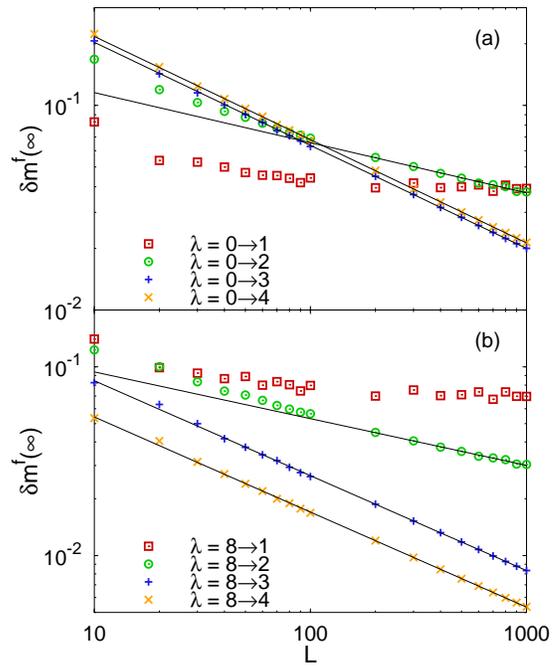}
	\caption{(Color online) Finite-size scaling of $\delta m^f(\infty)$ (see text) for the
quenches studied in Fig.~\ref{fig:fermdyn}. Thin continuous lines depict power-law fits in which
$\delta m^f(\infty)\propto L^{-0.25\pm0.01}$ for quenches with $\lambda_F=\lambda_c$ and
$\delta m^f(\infty)\propto L^{-0.500\pm0.005}$ in quenches with $\lambda_F>\lambda_c$. The power laws were
fitted to results with $L$ in the interval [200, 1000].}
	\label{fig:fermscal}
\end{figure}

To gain a quantitative understanding of the dependence of the long-time behavior of
Eq.~\eqref{eq:deltaO} on the system size, we consider the average of $\delta m^f(\tau)$ over the time
interval $\tau \in [10^5,10^6]$, and denote this quantity by $\delta m^f(\infty)$. We regard $\delta
m^f(\infty)$ as representative of the constant value about which $\delta m^f(\tau)$ fluctuates after
any transient dynamics have subsided, and away from any revival. In Fig.~\ref{fig:fermscal}, we plot
$\delta m^f(\infty)$ against $L$ for all the quenches shown in Fig.~\ref{fig:fermdyn}. The scalings
make apparent that $\delta m^f(\infty)$ converges to a finite value as the system size is increased
in quenches to the delocalized phase; i.e., although the infinite time average of each $m^f_k$
trivially agrees with its expectation value in the GGE, its instantaneous value does not relax to the
GGE in the thermodynamic limit.  By contrast, in quenches to the localized regime $\delta
m^f(\infty)$ decreases with increasing system size as a power law, which we find to be close to
$L^{-0.50}$. In the quenches to the critical regime, $\delta m^f(\infty)$ is much slower to reach a
clear power-law scaling, but for large system sizes its behavior is consistent with $\delta
m^f(\infty) \propto L^{-0.25}$.

The behavior of $\delta m^f(\infty)$ in the delocalized and localized regimes is exactly opposite to
that observed for the long-time average density difference \cite{NoteGramsch} in
Refs.~\cite{Gramsch2012}. There it was found that the normalized difference of the density $\delta
n(\infty)\propto L^{-0.5}$ in quenches to the delocalized phase, whereas it converges to a finite
value in quenches to the localized phase. Intuitively, the extended states in the delocalized regime
of the quasiperiodic lattice (and \emph{a fortiori} those in a monochromatic lattice) can be thought
of as states that are localized in momentum space. In fact, the correspondence between position
(momentum) space results in the delocalized phase and momentum (position) space quantities in the
localized phase can, in the case of free fermions, be seen to be an exact consequence of the
well-known self-duality of the Aubry-Andr\'e model~\cite{Sokoloff1985}. We note also that the
behavior of $\delta m^f(\infty)$ at the critical point is similar to that found for $\delta
n(\infty)$ in Ref.~\cite{Gramsch2012}.

Aside from the peculiar behavior observed at the critical point, which is understandable given the
very special character of the single-particle problem in this regime, it is remarkable that whenever
relaxation of $n_j$ or $m^f_k$ takes place (in the delocalized or localized regime)
the time fluctuations are proportional to $L^{-0.5}$.  This is consistent with the Gaussian
equilibration scenario of Ref.~\cite{campos_zanardi_13}, in which the square root of the normalized
time variance of one-body observables in noninteracting fermion systems was argued to scale as
$1/\sqrt{L}$~\cite{noteMeanabsolute}.

\begin{figure}[!t]
	\centering
	\includegraphics[width=0.475\textwidth]{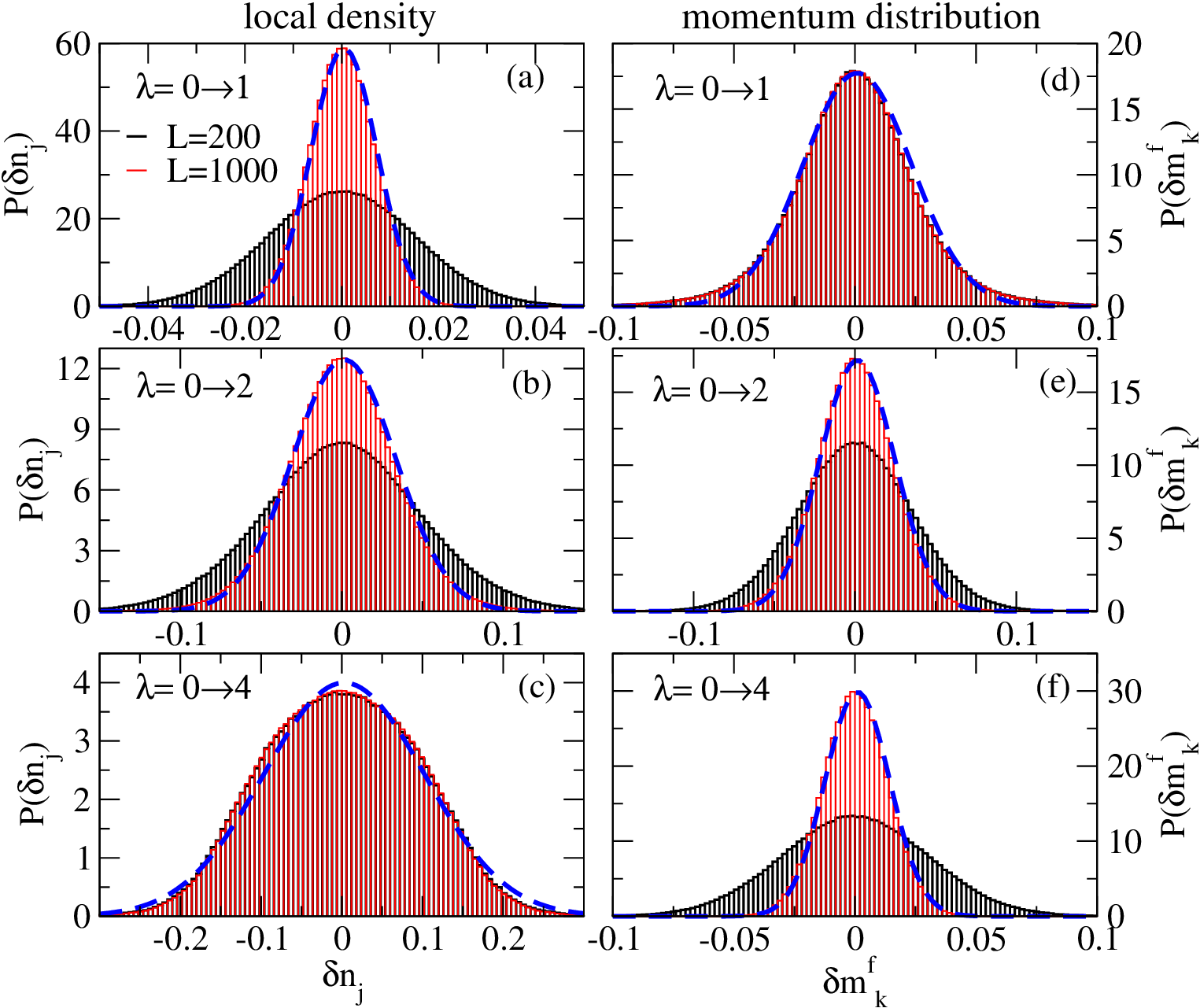}
  \vspace{-0.1cm}
	\caption{(Color online) Histograms of the time fluctuations of $n_{j}$ (left panels)
and $m^f_{k}$ (right panels) in quenches with $\lambda_I=0$ and $\lambda_F=1$ (a),(d),
$\lambda_F=2$ (b),(e), and $\lambda_F=4$ (c),(f). In all panels, we report results for systems
with $L=200$ and $L=1000$, and Gaussian fits to the data for $L=1000$ (dashed lines). From the fits, the means
were found to be zero ($\pm 0.003$) in all cases and the variances were found to be
$\sigma=(6.770\pm 0.004)\times10^{-3}$ (a), $(3.206\pm 0.009)\times10^{-2}$ (b),
$(1.008 \pm 0.007)\times10^{-1}$ (c), $(2.25 \pm 0.02)\times10^{-2}$ (d),
$(2.322 \pm 0.005)\times10^{-1}$ (e), and $(1.336 \pm 0.001)\times10^{-2}$ (f).}
	\label{fig:fermshist1}
\end{figure}

\begin{figure}[!t]
	\centering
	\includegraphics[width=0.475\textwidth]{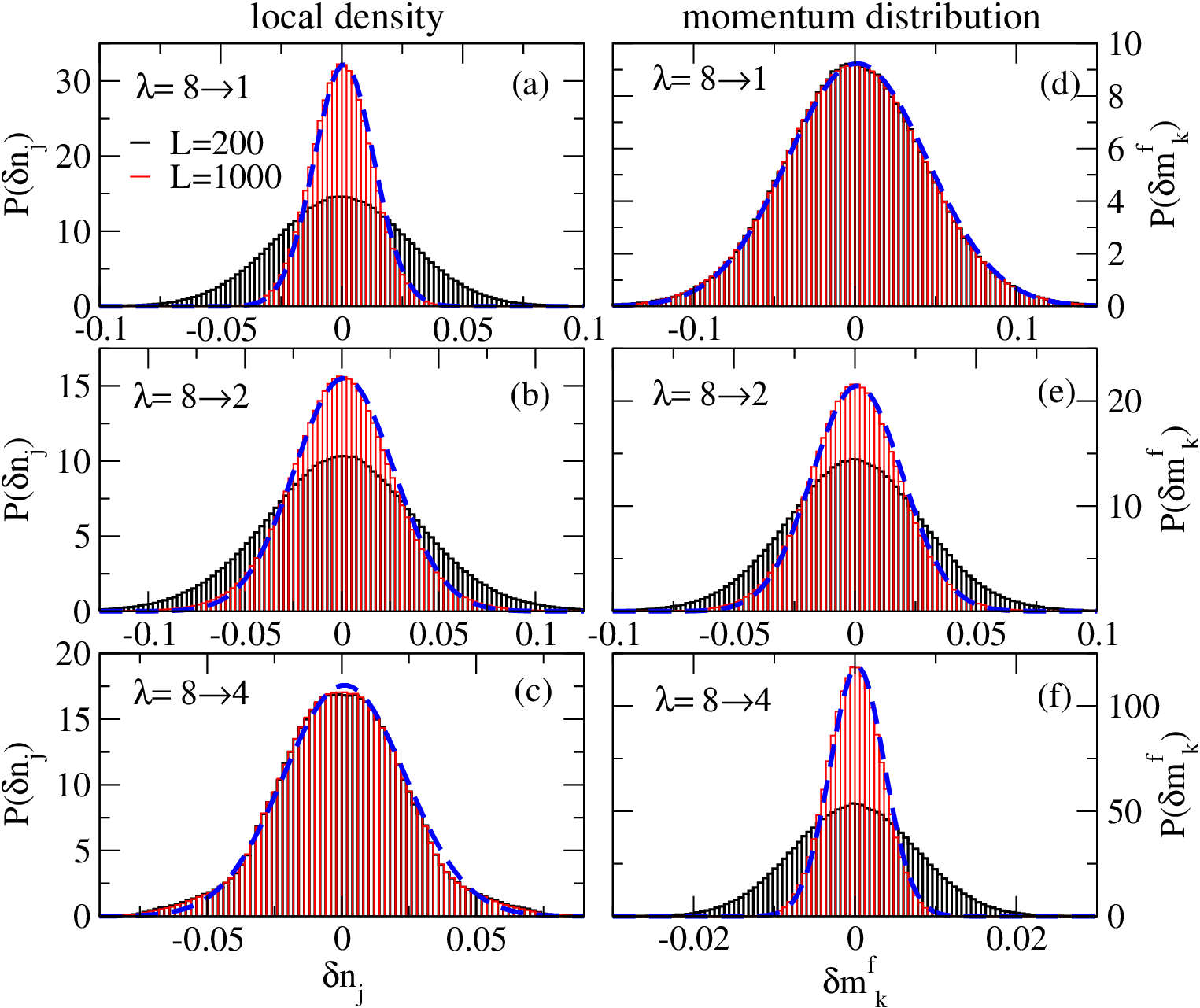}
  \vspace{-0.1cm}
	\caption{(Color online) Histograms of the time fluctuations of $n_{j}$ (left panels) and
$m^f_{k}$ (right panels) in quenches with $\lambda_I=8$ and $\lambda_F=1$ (a),(d), $\lambda_F=2$ (b),(e),
and $\lambda_F=4$ (c),(f). In all panels, we report results for systems with $L=200$ and $L=1000$, and Gaussian
fits to the data for $L=1000$ (dashed lines). From the fits, the means were found to
be zero ($\pm 0.003$) in all cases and the variances were found to be $\sigma=(1.2373\pm0.0004)\times10^{-2}$ (a),
$(2.570 \pm 0.008)\times10^{-2}$ (b), $(2.27 \pm0.01)\times 10^{-2}$ (c), $(4.318 \pm 0.005)\times10^{-2}$ (d),
$(1.860 \pm 0.004)\times10^{-2}$ (e), and $(3.360\pm 0.003)\times 10^{-3}$ (f).}
	\label{fig:fermhist2}
\end{figure}

In Figs.~\ref{fig:fermshist1} and \ref{fig:fermhist2}, we present histograms of the distributions of
differences $\delta n_j(\tau) = n_j(\tau) - \langle \hat{n}_j\rangle_\mathrm{GGE}$ between the
instantaneous values of the site occupations and their mean values in the GGE (left panels), and the
analogous quantities $\delta m^f_k(\tau) = m^f_k(\tau)-\langle \hat{m}^f_k\rangle_\mathrm{GGE}$
calculated for momentum-mode occupations (right panels). These histograms represent the full
distribution of fluctuations of the occupations of all lattice sites $j$, and all momentum modes $k$,
in quenches with $\lambda_I=0$ (Fig.~\ref{fig:fermshist1}) and $\lambda_I=8$
(Fig.~\ref{fig:fermhist2}). Once again, the results for a given value of $\lambda_F$ can be seen to
be qualitatively similar independently of the initial state. The histograms of the time fluctuations
of $n_j$ in quenches to the delocalized regime and of $m^f_k$ in quenches to the localized regime
have a Gaussian shape with a width that decreases with increasing system size. By contrast, the
histograms of the time fluctuations of $n_j$ in quenches to the localized regime and of $m^f_k$ in
quenches to the delocalized regime are in general non-Gaussian (as can be seen by comparing them to
the indicated best Gaussian fits to the distributions) and the widths of the distributions are not
seen to decrease with increasing system size. We note that the distributions of time fluctuations of
\emph{individual} lattice-site (momentum-mode) occupations in the localized (delocalized) phase,
which we have not shown, are quite strongly non-Gaussian, exhibiting, e.g., bimodal structures. The
behavior of the time fluctuations at the critical point is intermediate between what is seen in the
localized and delocalized phases: the fluctuations of $n_j$ and $m^f_k$ are close to Gaussian, with a
width that decreases less dramatically with increasing system size, consistent with the $L^{-0.25}$
behavior found for $\delta m^f(\infty)$ (Fig.~\ref{fig:fermscal}). This exotic behavior at the
critical point warrants a more specific investigation that is beyond the scope of this article.

An understanding of why Gaussian equilibration fails to occur for both fermionic observables at the
critical point, for $m^f_k$ in the delocalized phase, and for $n_j$ in the localized one, can be
gained through an analysis of the properties of the single-particle eigenstates of ${\hat H}_f$
($|s\rangle=\hat{\gamma}^\dagger_s|0\rangle$) in both real and momentum space
\cite{cazalilla_iucci_12,ziraldo_silva_12,ziraldo_santoro_13}. Since the variances
\begin{eqnarray}
\sigma^2_{n_j}&=&\lim_{\tau'\rightarrow \infty} \frac{1}{\tau'}
\int^{\tau'}_0 d\tau\, [n_j(\tau)-\langle \hat{n}_j\rangle_\textrm{GGE}]^2,\nonumber\\
 \sigma^2_{m_k}&=&\lim_{\tau'\rightarrow \infty} \frac{1}{\tau'}
\int^{\tau'}_0 d\tau\, [m^f_k(\tau)-\langle \hat{m}^f_k\rangle_\text{GGE}]^2,
\end{eqnarray}
can be written as
\begin{eqnarray}
 \sigma^2_{n_j}&=&\sum_{s\neq s'}|u_{js}|^2 |u_{js'}|^2 |\rho^I_{ss'}|^2, \nonumber\\
 \sigma^2_{m_k}&=&\sum_{s\neq s'}|v_{ks}|^2 |v_{ks'}|^2 |\rho^I_{ss'}|^2,
\end{eqnarray}
where  $\rho^I_{ss'}=\langle \Psi_I|\hat{\gamma}^\dagger_s \hat{\gamma}^{\phantom{\dagger}}_{s'}
|\Psi_I\rangle$, and $\hat{\gamma}^\dagger_s=\sum_j u_{js} \hat{f}^\dagger_j=\sum_k v_{ks}
\hat{c}^\dagger_k$ ($\hat{c}^\dagger_k$ creates a SF at momentum $k$), it follows that
\cite{ziraldo_silva_12}: (i) if $|s\rangle$ is delocalized in real (momentum) space then
$|u_{js}|^2\sim1/L$ ($|v_{ks}|^2\sim1/L$) and $\sigma^2_{n_j}$ ($\sigma^2_{m_k}$) must decrease as
$1/L$ or faster (because $\sum_{s,s'}|\rho^I_{ss'}|^2=N_f$) with increasing system size, and (ii) if
$|s\rangle$ is localized in real (momentum) space, then the corresponding variance $\sigma^2_{n_j}$
($\sigma^2_{m_k}$) will be dominated by rare large values of $|\rho^I_{ss'}|^2$ and will remain
finite in the thermodynamic limit.

\begin{figure}[!t]
	\centering
	\includegraphics[width=0.4\textwidth]{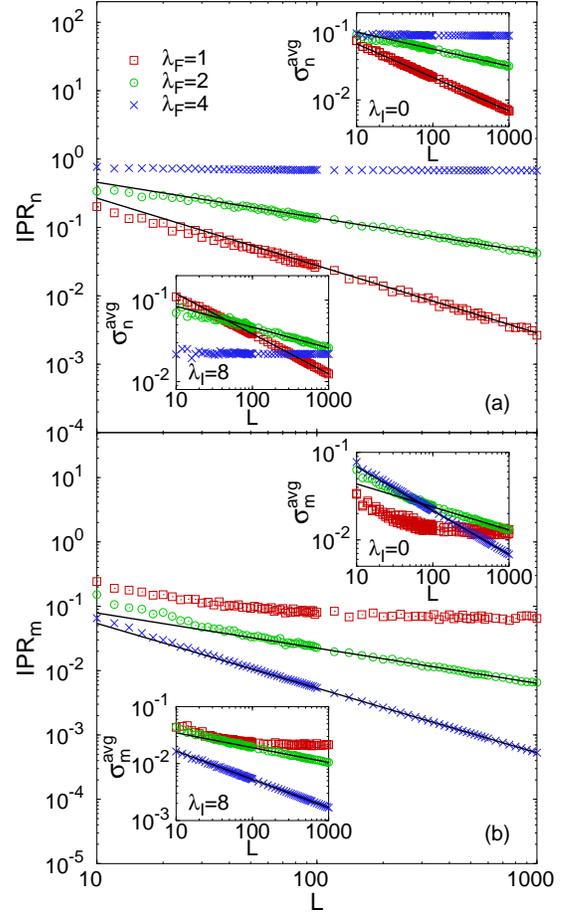}
	\caption{(Color online) Inverse participation ratios in real (a) and momentum space (b)
in the delocalized and localized phases as well as at the critical point. Thin continuous
lines are power-law fits in which (a) $\text{IPR}_n\propto L^{-0.99\pm0.02}$ for $\lambda_F=1$ and
$\text{IPR}_n\propto L^{-0.52\pm0.01}$ for $\lambda_F=2$, and (b) $\text{IPR}_m\propto
L^{-0.55\pm0.01}$ for $\lambda_F=2$ and $\text{IPR}_n\propto L^{-1.005\pm0.001}$ for $\lambda_F=4$.
The insets in (a) depict
$\sigma_n^\text{avg}$ for quenches with $\lambda_I=0$ (top inset) and $\lambda_I=8$ (bottom inset),
while the insets in (b) depict $\sigma_m^\text{avg}$ for quenches with $\lambda_I=0$ (top inset)
and $\lambda_I=8$ (bottom inset). Thin continuous lines are power-law fits in which, in both insets in
(a), $\sigma_n^\text{avg}\propto L^{-0.50\pm0.01}$ for $\lambda_F=1$ and
$\sigma_n^\text{avg}\propto L^{-0.25\pm0.01}$
for $\lambda_F=2$, and, in both insets in (b), $\sigma_m^\text{avg}\propto L^{-0.25\pm0.02}$ for
$\lambda_F=2$ and $\sigma_m^\text{avg}\propto L^{-0.500\pm0.005}$ for $\lambda_F=4$. The power laws were
fitted to results with $L$ in the interval [160, 1000].}
\label{fig:fermvar}
\end{figure}

In the insets to Fig.~\ref{fig:fermvar}, we plot the average of the square root of the variance over
all lattice sites $\sigma_n^\text{avg}=1/L\sum_j \sigma_{n_j}$ [insets in Fig.~\ref{fig:fermvar}(a)]
and over all momentum states $\sigma_m^\text{avg}=1/L\sum_k\sigma_{m_k}$ [insets in
Fig.~\ref{fig:fermvar}(b)] against the system size $L$, for systems with between 10 and 1000 lattice
sites and subject to the same quenches studied previously ($\lambda_I=0$ in the top insets and
$\lambda_I=8$ in the bottom ones). The results for $\sigma_n^\text{avg}$ and $\sigma_m^\text{avg}$
can be seen to be qualitatively similar to those in Ref.~\cite{Gramsch2012} for $\delta n(\infty)$
and in Fig.~\ref{fig:fermscal} for $\delta m^f(\infty)$. In order to relate the behavior of the
variances to the properties of the single-particle eigenstates $|s\rangle$ in real and momentum
space, we compute the average inverse participation ratios (IPRs)
\begin{eqnarray}
 \text{IPR}_n&=&\frac{1}{L}\sum_s\sum_j |u_{js}|^4,\nonumber\\
 \text{IPR}_m&=&\frac{1}{L}\sum_s\sum_k |v_{ks}|^4.
\end{eqnarray}
The results for the IPRs are reported in the main panels in Fig.~\ref{fig:fermvar}. They show that,
as expected, the eigenstates of the Hamiltonian are delocalized in real space ($\text{IPR}_n\sim1/L$)
and localized in momentum space ($\text{IPR}_m\sim L^0$) for $\lambda_F<2$, and localized in real
space ($\text{IPR}_n\sim L^0$) and delocalized in momentum space ($\text{IPR}_m\sim1/L$) for
$\lambda_F>2$, which explains the behavior of the time fluctuations of the density and fermionic
momentum distributions in the delocalized and localized regimes. For $\lambda_F=2$, we find that
$\text{IPR}_n\sim\text{IPR}_m\sim 1/\sqrt{L}$, illustrating the exotic structure of the eigenstates
of the critical Hamiltonian in both real and momentum space. From this scaling we can infer that
$|u_{js}|^2\sim|v_{ks}|^2\sim1/L^{0.75}$ when $\lambda_F=2$, implying that $\sigma_{n_j}^2$ and
$\sigma_{m_k}^2$ decay like $1/\sqrt{L}$ or faster at the critical point, which is indeed consistent
with the results of Fig.~\ref{fig:fermscal}. We note also that in general, the average square-root
variances and IPRs in the critical regime reach their exotic scaling limits at quite small values of
$L$, compared to the behavior of $\delta m^f(\infty)$ at the critical point
(Fig.~\ref{fig:fermscal}). The unambiguous scaling at criticality seen in Fig.~\ref{fig:fermvar} thus
lends strong support to our identification of $\delta m^f(\infty)$ as scaling like ${\sim}L^{-0.25}$
in this regime.

For fermionic observables, we should stress the fact that, in the localized regime, localization in
real space precludes relaxation of the density profiles in the same way that, in the delocalized regime,
localization in momentum space precludes relaxation of the momentum distribution. This symmetry between
localization in real and momentum space is broken in the case of HCBs, because the mapping to the
underlying model of SFs only preserves correlations that are diagonal in real space (i.e., properties related
to the density). Thus for HCBs, although relaxation of the density is precluded in the localized
regime, relaxation of the momentum distribution can occur in the delocalized regime, as was found in
Ref.~\cite{Gramsch2012}.

As discussed in Refs.~\cite{cazalilla_iucci_12,ziraldo_silva_12,ziraldo_santoro_13}, if the variances
of one-particle correlations (here we have focused only on the density and momentum distributions) do
not vanish with increasing system size---which is only possible if off-diagonal elements of
$|\rho^I_{ss'}|^2$ contribute to the fluctuations in the thermodynamic limit---then Wick's theorem
can break down for time averages of higher-order correlations. We recall that nonlocal one-particle
correlations of HCBs [e.g., Eq.~\eqref{momBosons}] correspond, via the Jordan-Wigner transformation,
to higher-order correlations of SFs~\cite{ziraldo_santoro_13}. Thus the breakdown of the GGE for
describing $m^b_k$ after relaxation in the localized phase of HCBs \cite{Gramsch2012} can be
understood as a direct consequence of the fact that time fluctuations of local one-particle
correlations of the underlying free-fermion model remain finite as one increases the system size in
the localized phase.

\section{Many-Body Analysis}
\label{Sec:hcbs}

In this section, we study the density profiles and momentum distribution functions in the many-body
eigenstates of the SF and HCB Hamiltonians. Our goal is to understand how localization, in real and
momentum space, affects the many-body eigenstate expectation values of these observables. Since this
study requires the construction of the full set of energy eigenstates of the many-body system, which
grows exponentially with increasing system size, our analysis will be restricted to lattice lengths
that are much smaller than those studied in the single-particle analysis of the previous section. The
smallest systems considered in this section have 20 sites and the largest ones have $L=50$, and in
each case $N=L/5$, in contrast to the filling $N=L/2$ (which yields the maximal Hilbert space
dimension for a given system size) considered in the previous section. The largest systems we
consider here have a Hilbert space of dimension $O(10^{10})$. In order to compare systems that have
equivalent excitation energies per particle after the quench, we will focus on three quenches that,
while having the same final Hamiltonians as in the previous section, have for their initial states
the respective ground states of Hamiltonians with $\lambda_I=-1.5$ for $\lambda_F=1.0$,
$\lambda_I=-0.5$ for $\lambda_F=2.0$, and $\lambda_I=1.0$ for $\lambda_F=4.0$. These three quenches
lead to time-evolving states whose energies are similar to those of systems in thermal equilibrium
with $T\sim 1.7$.

\begin{figure}[t!]
\centering
\includegraphics[width=0.475\textwidth, clip]{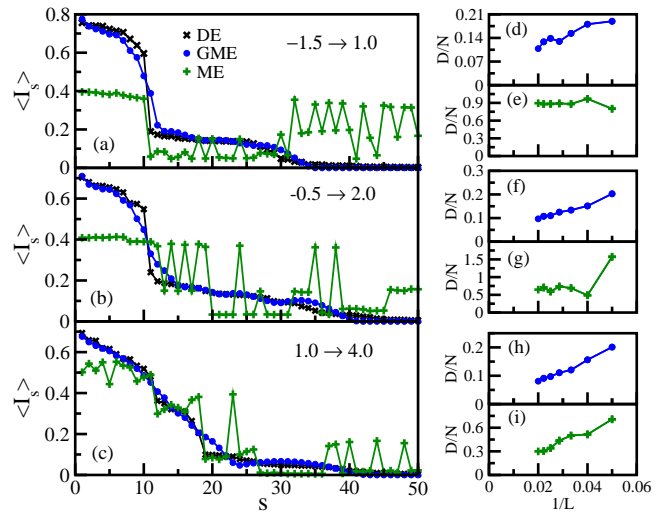}
\caption{(Color online) [(a),(b),(c)] Conserved quantities $\langle \hat I_s \rangle$ in the DE, GME, and ME
for $L=50$, $N=10$, sorted in order of decreasing occupation in the DE. [(d)--(i)] Scaling
of the absolute differences $D$ between distributions of conserved quantities with increasing system size.
[(d),(f),(h)] Absolute differences $D$ between
the conserved quantities in the GME and the DE [see Eq.~\eqref{eq:madD}]. [(e),(g),(i)] $D$ between the
conserved quantities in the ME and the DE [same as Eq.~\eqref{eq:madD} but with ``GME''$\rightarrow$``ME''].
The quenches are indicated as $\lambda_I \rightarrow \lambda_F$. For quenches
$\lambda_I=-1.5 \rightarrow \lambda_F=1.0$ and $\lambda_I=-0.5 \rightarrow\lambda_F=2.0$ we found that
$\delta_\textrm{GME}=0.85$ yields the minimum value of $D$, and for quenches
$\lambda_I=1.0 \rightarrow \lambda_F=4.0$ we found $\delta_\textrm{GME}=0.95$. For the ME,
$\delta_\textrm{ME}=0.05$ in all quenches.}
\label{fig:consquant}
\end{figure}

In Fig.~\ref{fig:consquant}, we show the distribution of conserved quantities in the quenches
described above for systems with $L=50$. We compare results for those distributions in the initial
state (same as the DE and the GGE), in the GME (constructed as described in Sec.~\ref{Sec:model}),
and in the ME. The contrast between the distribution of conserved quantities in the initial state and
in the ME is apparent, whereas the distribution in the GME closely agrees with that in the initial
state. Figures \ref{fig:consquant}(d)--\ref{fig:consquant}(i) depict how the normalized absolute
differences between the distribution of conserved quantities in the GME and ME and that in the
initial state [Eq.~\eqref{eq:madD} and the analogous expression obtained by replacing $\langle
\hat{I}_s\rangle_\mathrm{GME}$ with $\langle \hat{I}_s\rangle_\mathrm{ME}$, respectively] scale with
increasing system size. For the systems analyzed here, these differences are always smaller for the
GME than for the ME. More importantly, for all quenches, they are seen to decrease with increasing
system size for the GME [Figs.~\ref{fig:consquant}(d), \ref{fig:consquant}(f), and
\ref{fig:consquant}(h)]. The differences between the ME and the DE exhibit clear saturation behavior
in the delocalized and critical regimes [Figs.~\ref{fig:consquant}(e) and \ref{fig:consquant}(g)]. In
the localized regime [Figs.~\ref{fig:consquant}(i)], the difference between the ME and DE is both
smaller than, and does not exhibit saturation quite as obviously as, that in the other two regimes.
We note, however, that it was found previously in Ref.~\cite{Cassidy2011} that the discrepancy
between distributions of conserved quantities in the ME and DE can be strongly dependent on the
initial state (and in particular on its energy).

Although disagreement between the distributions of conserved quantities in the DE and ME
indicates the failure of the ME to describe the state of the system after relaxation, the degree of agreement
between the corresponding distributions in the DE and GME yields only incomplete information on the accuracy of the
GGE as a characterization of the system at long times.  In particular, the distributions here apply equally to SFs and
HCBs, whereas the results of Ref.~\cite{Gramsch2012} and Sec.~\ref{Sec:fermions} indicate that the presence or absence
of relaxation, and the agreement between time averages and the GGE predictions, can differ between SFs and HCBs
with equal quench parameters.  To further characterize the relationship between (generalized) thermalization
and the structure of the many-body eigenstates of the system, we now turn our attention to the behavior of
the density and momentum distributions in the eigenstates of the SF and HCB systems.

In order to quantify the differences between the predictions of the GME and ME for each observable
and those of the DE, we compute the normalized difference,
\begin{equation}
\Delta O^{\textrm{stat-DE}} = \frac{\sum_\ell |O_{\textrm{stat},\ell}  - O_{\textrm{DE},\ell}|}
{\sum_\ell O_{\textrm{DE},\ell}},
\label{eq:Odiff}
\end{equation}
where $O$ is either the site occupation $n$ (for which $\ell=j$) or the momentum occupation $m^f$ or
$m^b$ (for which $\ell=k$), $O_{\textrm{stat},\ell}=\sum_\alpha
w^{\textrm{stat}}_\alpha\langle\psi_\alpha|\hat{O}_\ell|\psi_\alpha\rangle
\equiv\langle\hat{O}_\ell\rangle_\textrm{stat}$, with $w^{\textrm{stat}}_\alpha$ the weight of each
many-body eigenstate in the relevant ensemble, and ``stat'' stands for one of the three ensembles:
the DE, the GME, or the ME. Hence, the agreement between a statistical ensemble and the DE in the
thermodynamic limit becomes apparent if $\Delta O^\textrm{stat-DE}$ vanishes with increasing system
size.

We quantify the behavior of the eigenstate expectation values of the observables by calculating the
average variance within each ensemble, which we define by
\begin{equation}
\sigma_O^{\textrm{stat}} = \frac{\sum_\ell  \sqrt{ (O^2)_{\textrm{stat},\ell} - (O_{\textrm{stat},\ell})^2}}{L},
\end{equation}
where $(O^2)_{\textrm{stat},\ell}=\sum_\alpha w^{\textrm{stat}}_\alpha \langle\psi_\alpha
|\hat{O}_\ell|\psi_\alpha\rangle^2\neq \langle\hat{O}^2_\ell\rangle_\textrm{stat}$ (cf.
Refs.~\cite{rigol09STATa,rigol09STATb}). A finite value of ($\sigma_O^{\textrm{GME}}$)
$\sigma_O^{\textrm{ME}}$ in the thermodynamic limit implies that (generalized) eigenstate
thermalization does not occur. However, the vanishing of ($\sigma_O^{\textrm{GME}}$)
$\sigma_O^{\textrm{ME}}$ is not a sufficient condition for (generalized) eigenstate thermalization to
occur: the (GME) ME may still fail to describe observables after relaxation if the differences
between the expectation values of observables in distinct eigenstates contributing to the ensemble do
not vanish in the thermodynamic limit. In such a scenario, the results of the DE for the expectation
values of observables may be dominated by so-called ``rare'' states that exhibit expectation values
for the observables that are significantly different from the (GME) ME averages, but constitute a
sufficiently small proportion of all states in the ensemble that they do not preclude
($\sigma_O^{\textrm{GME}}$) $\sigma_O^{\textrm{ME}}$ from vanishing with increasing system size
\cite{biroli_kollath_10}. To investigate this possibility, we calculated the individual differences
between observables in each eigenstate and in the ensemble average and attempted to quantify the
scaling of the maximal differences with increasing system size. However, the results were found to be
dominated by finite-size effects and we could not extract any consistent information from our
investigations. We therefore only report results for $\sigma_O^{\textrm{stat}}$ in this section.

We further note that, in a previous study of HCBs systems whose properties are qualitatively
similar to those of the systems studied here for $\lambda<\lambda_c$ \cite{Cassidy2011},
indications were found that $\sigma_m^{\textrm{GME}}$ (for ${m}^b_k$) vanishes with increasing system size
while $\langle\hat{m}^b_k\rangle_\textrm{GME}=\langle\hat{m}^b_k\rangle_\textrm{DE}$. Since the weights
$w^{\textrm{GME}}_\alpha$ and $w^{\textrm{DE}}_\alpha$ were seen to be different \cite{Cassidy2011},
with an exponentially smaller number of states usually contributing to the DE when compared to the GGE
\cite{Santos2011PRL}, the results of Ref.~\cite{Cassidy2011} hint that a generalized eigenstate thermalization
is at play for $m^b_k$. Namely, that all eigenstates with similar distributions of conserved quantities
have similar expectation values of the HCB momentum distribution function (with differences that vanish
in the thermodynamic limit).

\begin{figure}[tb]
\centering
\includegraphics[width=0.475\textwidth]{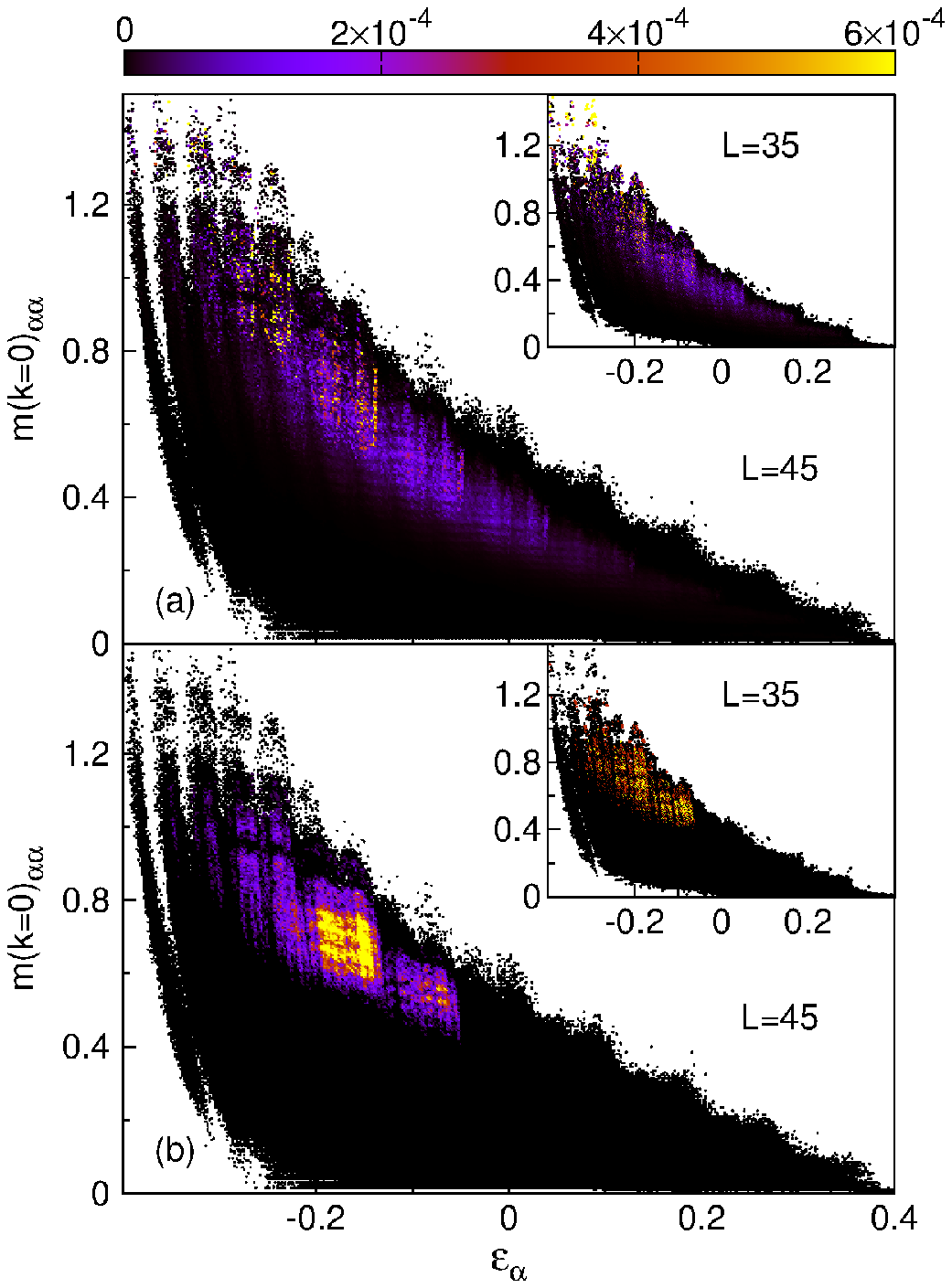}
\caption{(Color online) Density plots of the coarse-grained weights of energy eigenstates in (a) the DE and (b) the GME
as a function of the expectation value of $m(k=0)$ and the eigenstate energy per site $\varepsilon_\alpha=E_\alpha/L$,
for quenches of HCBs to the delocalized regime. Results are shown for systems with $L=45$ (main panel) and $L=35$ (insets).
The width of the windows used in the coarse graining is $5\times10^{-3}$ in $m(k=0)$ and $2\times10^{-3}$ in $\varepsilon$.}
\label{fig:Fig07_DensPlotD}
\end{figure}

We now present results that indicate that this mechanism is indeed the explanation why the GGE
provides an accurate description of HCB observables after relaxation in quenches to the delocalized
phase. In Fig.~\ref{fig:Fig07_DensPlotD}, we show density plots of the coarse-grained weights with
which eigenstates contribute to the DE [Fig.~\ref{fig:Fig07_DensPlotD}(a)] and to the GME
[Fig.~\ref{fig:Fig07_DensPlotD}(b)] for delocalized HCB systems with $L=45$ (main panels) and $L=35$
(insets). We see that in both ensembles, as the system size increases, weight becomes increasingly
concentrated in eigenstates with $-0.2<\varepsilon_\alpha\equiv E_\alpha/L<-0.1$ (though more clearly
in the GME than in the DE). Moreover, the expectation values of $\hat{m}(k=0)$ in these most highly
weighted eigenstates are narrowly distributed compared to the full range of expectation values of all
eigenstates within the same energy range. Furthermore, it is remarkable that the expectation values
of $\hat{m}(k=0)$ in the dominant states of the two ensembles are similar to each other, and that
this agreement is seen to improve with increasing system size.

\begin{figure}[tb]
\centering
\includegraphics[width=0.47\textwidth]{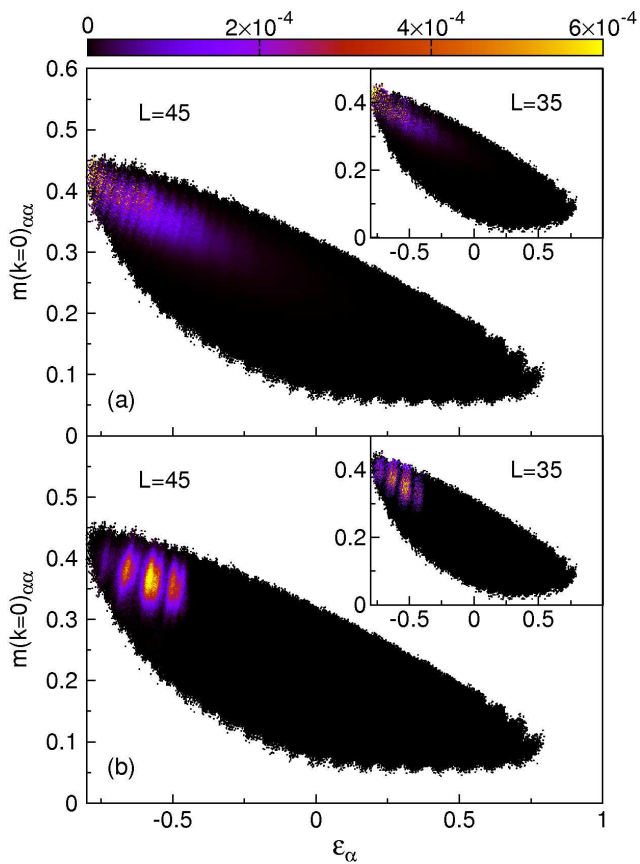}
\caption{(Color online) Density plots of the coarse-grained weights of energy eigenstates in (a) the DE and (b) the GME
as a function of $m(k=0)$ and the eigenstate energy per site $\varepsilon_\alpha=E_\alpha/L$, for quenches
of HCBs to the localized regime. Results are shown for systems with $L=45$ (main panel) and $L=35$ (insets).
The width of the windows used in the coarse graining is $10^{-3}$ in $m(k=0)$ and $4\times10^{-3}$ in $\varepsilon$.}
\label{fig:Fig08_DensPlotL}
\end{figure}

In Fig.~\ref{fig:Fig08_DensPlotL} we present the corresponding results for quenches to the localized phase.
We observe qualitative differences between the distributions of weights in the DE and in the GME:
the former are spread relatively smoothly over energies $\varepsilon_\alpha<-0.3$ [Fig.~\ref{fig:Fig08_DensPlotL}(a)],
whereas the latter tend to concentrate in several distinct energy bands [Fig.~\ref{fig:Fig08_DensPlotL}(b)].
Moreover, in the DE the weights tend to increase as $\varepsilon_\alpha$ decreases, corresponding to larger values of
the expectation of $\hat{m}(k=0)$, whereas in the GME the weight distribution is more strongly concentrated in
eigenstates with higher values of $\varepsilon_\alpha$, and smaller values of the zero-momentum expectation values.
This suggests that the dominant eigenstates in each of the two ensembles do not
yield similar momentum distribution functions, implying the failure of generalized eigenstate thermalization in
the localized phase, similar to the failure of eigenstate thermalization in nonintegrable systems in the presence
of localization~\cite{khatami_12}.

In what follows, we explore in more detail the properties of the eigenstate expectation values of particular
observables of interest for SFs and HCBs, and their dependence on the system size.

\subsection{Quenches to the delocalized regime}

We start by studying the behavior of the density profiles in the many-body eigenstates of the SF and
HCB Hamiltonians. The density profiles of SFs and HCBs are identical, and, by construction, the
predictions of the GGE for the expectation value of this observable are the same as those of the DE.

\begin{figure}[!b]
\centering
\includegraphics[width=0.475\textwidth, clip]{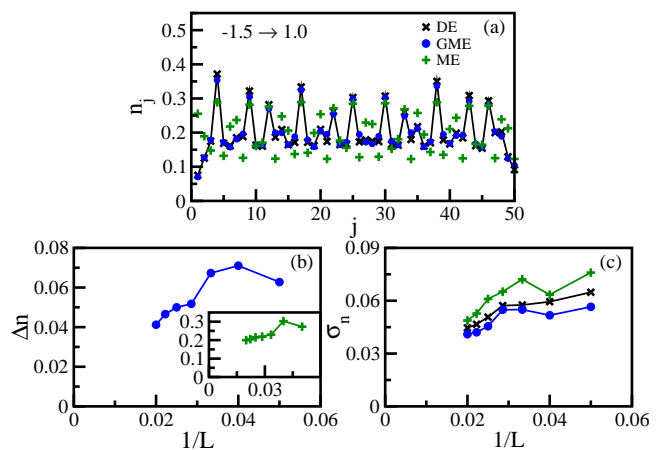}
\caption{(Color online)  (a) Density profiles ($L=50$), (b) $\Delta n^{\textrm{stat-DE}}$, and
(c) $\sigma_n^{\textrm{stat}}$ in quenches to the delocalized regime, $\lambda_I =-1.5\rightarrow\lambda_F=1.0$.
In (b), the main panel depicts $\Delta n^{\textrm{GME-DE}}$, while the inset depicts
$\Delta n^{\textrm{ME-DE}}$. The legends for the ensembles considered are the same in all panels and
are reported in (a). The values of $\delta_\textrm{GME}$ and $\delta_\textrm{ME}$ are the
same for all $L$'s and are equal to the ones in Fig.~\ref{fig:consquant}.}
\label{fig:denDeloc}
\end{figure}

In Fig.~\ref{fig:denDeloc}(a), we show the density profiles after a quench to the delocalized
regime as predicted by the DE, the GME, and the ME. The agreement between DE and GME is excellent, as evidenced by the
small values of $\Delta n^{\textrm{GME-DE}}$ in Fig.~\ref{fig:denDeloc}(b). The latter quantity is seen
to decrease with increasing system size indicating that the DE and GME predictions will agree in the thermodynamic
limit. On the other hand, in Fig.~\ref{fig:denDeloc}(a), large differences can be seen between the outcomes of DE
and ME calculations for the density profiles, which leads to large values of $\Delta n^{\textrm{ME-DE}}$
as depicted in the inset in Fig.~\ref{fig:denDeloc}(b). The results in the inset suggest that
$\Delta n^{\textrm{ME-DE}}$ saturates to a finite value with increasing $L$, which would be consistent with the
results of Ref.~\cite{Gramsch2012}, where it was shown that the outcomes of the relaxation dynamics for this observable
failed to approach the predictions of the grand-canonical ensemble with increasing system size (for systems up to 20 times larger
than those considered here).

Figure~\ref{fig:denDeloc}(c) shows the scaling of the average variance of the site occupations in the
DE, the GME, and the ME with $L$. In all ensembles, this variance is of a similar small magnitude, and
decreases with increasing system size. However, our results are not conclusive as to whether the
variance (in any of the ensembles) vanishes or  saturates to a finite value in the
thermodynamic limit.  Inasmuch as we understand thermalization in an isolated system to result from
eigenstate thermalization, the fact that this observable does not thermalize~\cite{Gramsch2012}
implies that eigenstate thermalization does not occur in this system. Given that
the predictions of the GGE and GME for $n_j$ agree with those of the DE,
it remains to be clarified in future studies whether the generalized eigenstate thermalization
scenario is valid for this observable or not.

\begin{figure}[!t]
\centering
\includegraphics[width=0.475\textwidth, clip]{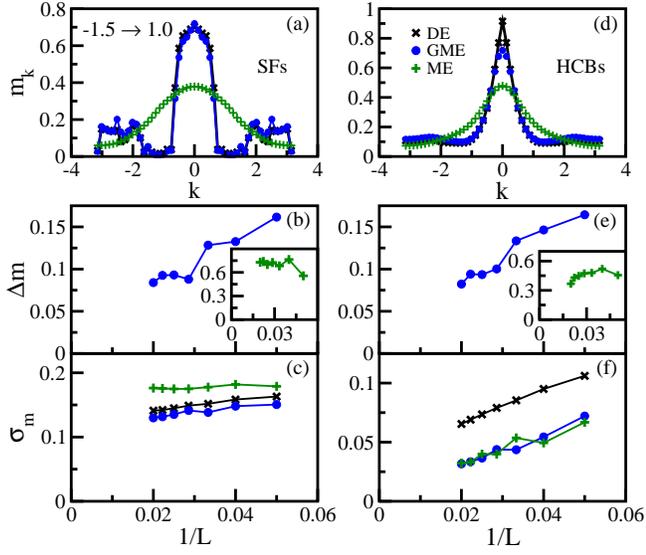}
\caption{(Color online)  [(a),(d)] Momentum distribution functions ($L=50$),
[(b),(e)] $\Delta m^{\textrm{stat-DE}}$, and [(c),(f)] $\sigma_m^{\textrm{stat}}$, for SFs (left column)
and HCBs (right column), in quenches to the delocalized regime, $\lambda_I =-1.5\rightarrow\lambda_F=1.0$.
In [(b),(e)], the main panels depict $\Delta m^{\textrm{GME-DE}}$, while the insets depict
$\Delta m^{\textrm{ME-DE}}$. The legends for the ensembles considered are the same in all panels and
are reported in (d). The values of $\delta_\textrm{GME}$ and $\delta_\textrm{ME}$ are the
same for all $L$'s and are equal to the ones in Fig.~\ref{fig:consquant}.}
\label{fig:momDeloc}
\end{figure}

In Fig.~\ref{fig:momDeloc}, we present a study equivalent to the one in Fig.~\ref{fig:denDeloc}, but
for the momentum distributions of SFs and HCBs. The first feature that is apparent in
Figs.~\ref{fig:momDeloc}(a) and \ref{fig:momDeloc}(d) is the contrast between the shapes of the
momentum distributions of SFs and HCBs. For small values of $k$, the former resembles a Fermi sea and
the latter resembles a bosonic system, both systems at finite but low temperature \cite{rigol_05}.
However, the high occupation of $m^f_k$ and $m^b_k$ in the tails makes it evident that the systems
are not in thermal equilibrium, as can be seen by comparing them to the predictions of the ME. By
contrast, the GME predictions for the momentum distributions closely agree with the DE results. This
can be seen more clearly in Figs.~\ref{fig:momDeloc}(b) and \ref{fig:momDeloc}(e), which show that
the differences between the GME and the DE predictions are small and decrease with increasing system
size. The insets to the same panels show that $\Delta m^{\textrm{ME-DE}}$ is several times larger
than $\Delta m^{\textrm{GME-DE}}$ for the system sizes studied, and that the fermionic $\Delta
m^{\textrm{ME-DE}}$ has a tendency to saturate to a finite value as $L\rightarrow\infty$, though the
behavior of $\Delta m^{\textrm{ME-DE}}$ for HCBs is less clear.

The results for the scaling of $\sigma_m^{\textrm{stat}}$ with increasing system size
[Figs.~\ref{fig:momDeloc}(c) and \ref{fig:momDeloc}(f)] make apparent a fundamental difference
between the behavior of eigenstate expectation values of $\hat{m}^f_k$ and $\hat{m}^b_k$. In each
ensemble, $\sigma_m^{\textrm{stat}}$ for SFs [Fig.~\ref{fig:momDeloc}(c)] is seen to be larger than
the corresponding variance $\sigma_m^{\textrm{stat}}$ for HCBs [Fig.~\ref{fig:momDeloc}(f)], and the
former appears to saturate to a finite value (more obviously in the ME than in the DE or GME), while
the latter appears to vanish (more clearly in the GME and ME than in the DE), in the thermodynamic
limit. This makes evident that localization of the single-particle fermionic eigenstates in momentum
space has a clear consequence on the behavior of $m^f_k$ in the many-body eigenstates, for which
$\sigma_m^{\textrm{stat}}$ may be finite in the thermodynamic limit. By contrast, such a localization
phenomenon in the single-particle basis of the SF model appears not to have any effect on the
many-body eigenstate expectation values of the momentum distribution of the corresponding HCBs, for
which {\it generalized} eigenstate thermalization may take place as $\sigma_m^{\textrm{stat}}$
appears to vanish for all ensembles, and
$\langle\hat{m}^b_k\rangle_\textrm{DE}=\langle\hat{m}^b_k\rangle_\textrm{GME}$ \cite{Gramsch2012}.
Again, we stress that even if the bosonic $\sigma_m^{\textrm{ME}}$ does vanish in the thermodynamic
limit, we can infer that eigenstate thermalization does not occur for $m^b_k$ from the failure of
this observable to thermalize~\cite{Gramsch2012}.

\subsection{Quenches to the localized regime}

\begin{figure}[!b]
\centering
\includegraphics[width=0.475\textwidth, clip]{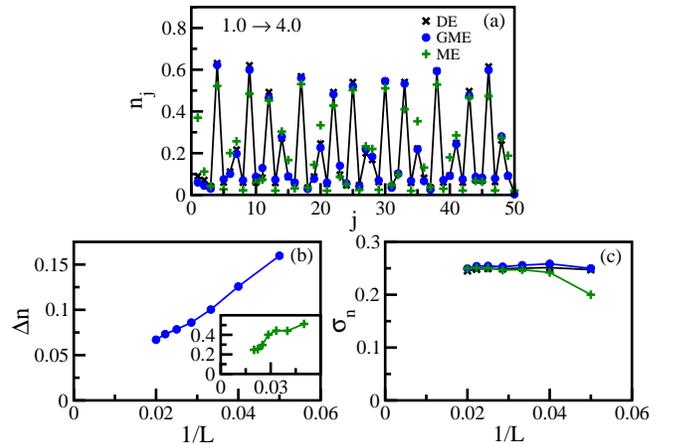}
\caption{(Color online)  (a) Density profiles ($L=50$), (b) $\Delta n^{\textrm{stat-DE}}$, and
(c) $\sigma_n^{\textrm{stat}}$ in quenches to the localized regime, $\lambda_I =1.0\rightarrow\lambda_F=4.0$.
In (b), the main panel depicts $\Delta n^{\textrm{GME-DE}}$, while the inset depicts
$\Delta n^{\textrm{ME-DE}}$. The legends for the ensembles considered are the same in all panels and
are reported in (a). The values of $\delta_\textrm{GME}$ and $\delta_\textrm{ME}$ are the
same for all $L$'s and are equal to the ones in Fig.~\ref{fig:consquant}.}
\label{fig:denLoc}
\end{figure}

Density profiles obtained within the DE, GME, and ME after a quench to the localized regime are shown
in Fig.~\ref{fig:denLoc}(a), and the scalings of $\Delta n^{\textrm{GME-DE}}$
($\Delta n^{\textrm{ME-DE}}$) with increasing system size are reported in the main panel
(inset) in Fig.~\ref{fig:denLoc}(b). Despite the fact that the site occupations fluctuate from site
to site much more in Fig.~\ref{fig:denLoc}(a) than in Fig.~\ref{fig:denDeloc}(a), the GME results still
closely agree with those of the DE, while the ME results do not. The scaling of
$\Delta n^{\textrm{GME-DE}}$ with increasing system size suggests that the differences
between the predictions of the GME and the DE will vanish in the thermodynamic limit.
The results for $\Delta n^{\textrm{ME-DE}}$ are less conclusive, although the values of this
quantity obtained for the largest system sizes suggest a possibly tendency toward saturation.

A clear difference between the behavior of the site occupations in the localized and
delocalized regimes is seen in the fact that the variance of the eigenstate
expectation values of this observable saturates in all three ensembles to a finite value in the former
[Fig.~\ref{fig:denLoc}(c)], whereas our results suggest that it vanishes in the latter
[Fig.~\ref{fig:denDeloc}(c)], as the system size is increased. The saturation observed in
Fig.~\ref{fig:denLoc}(c) is a clear consequence of localization of the single-particle eigenstates in
real space for $\lambda>\lambda_c$. Finite values of the fermionic $\sigma_m^{\textrm{ME}}$ in the
delocalized regime and of $\sigma_n^{\textrm{ME}}$ in the localized one are physically relevant examples of
the failure of the variance of few-body observables in the many-body eigenstates that constitute the ME to
vanish in the thermodynamic limit, contrary to what is generally expected to occur for few-body
observables \cite{biroli_kollath_10}.

\begin{figure}[t]
\centering
\includegraphics[width=0.475\textwidth, clip]{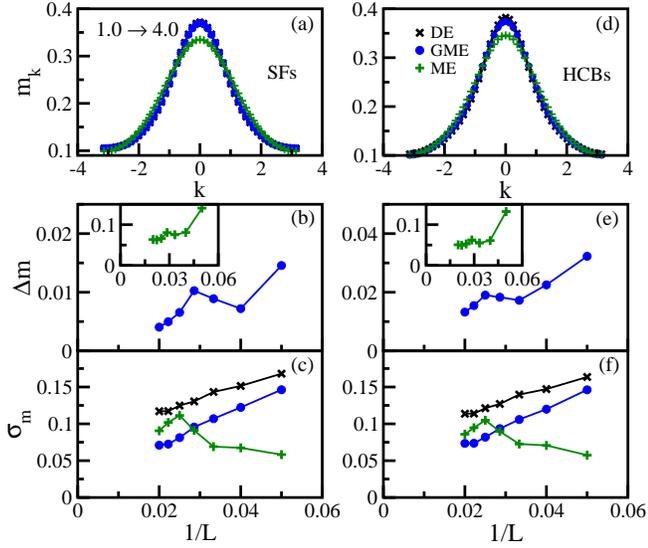}
\caption{(Color online) [(a),(d)] Momentum distribution functions ($L=50$),
[(b),(e)] $\Delta m^{\textrm{stat-DE}}$, and [(c),(f)] $\sigma_m^{\textrm{stat}}$, for SFs (left column)
and HCBs (right column), in quenches to the localized regime, $\lambda_I =1.0\rightarrow\lambda_F=4.0$.
In [(b),(e)], the main panels depict $\Delta m^{\textrm{GME-DE}}$, while the insets depict
$\Delta m^{\textrm{ME-DE}}$. The legends for the ensembles considered are the same in all panels and
are reported in (d). The values of $\delta_\textrm{GME}$ and $\delta_\textrm{ME}$ are the
same for all $L$'s and are equal to the ones in Fig.~\ref{fig:consquant}.}
\label{fig:momLoc}
\end{figure}

In Figs.~\ref{fig:momLoc}(a) and \ref{fig:momLoc}(d), we show the momentum distribution functions of SFs and
HCBs in the DE, the GME, and the ME after a quench to the localized regime. In each ensemble, the results for SFs
and HCBs are barely distinguishable from each other, which we might intuitively attribute to localization
undermining the particle statistics. As in the quenches to the delocalized regime, the GME results closely
follow those of the DE, and the main panels in Figs.~\ref{fig:momLoc}(b) and \ref{fig:momLoc}(e) indicate that
$\Delta m^{\textrm{GME-DE}}$ decreases with increasing system size. Whereas we expect $\Delta m^{\textrm{GME-DE}}$
to vanish for SFs in the thermodynamic limit (as the DE and GGE predictions coincide for all one-body
fermionic observables), for HCBs we expect it to converge to a small but finite value,
because of the failure of the GGE to describe the time averages of the HCB momentum distribution function observed
in Ref.~\cite{Gramsch2012}. In Figs.~\ref{fig:momLoc}(a) and \ref{fig:momLoc}(d), the ME results for the
momentum distributions are clearly distinct from those of the DE, and that difference is expected to remain
in the thermodynamic limit both for SFs and HCBs, as suggested by the scaling of $\Delta m^{\textrm{ME-DE}}$
in the insets in Figs.~\ref{fig:momLoc}(b) and \ref{fig:momLoc}(e).

Results for the scaling of $\sigma_m^{\textrm{stat}}$ with increasing system size are presented in
Figs.~\ref{fig:momLoc}(c) and \ref{fig:momLoc}(f), for SFs and HCBs, respectively.  Once again, the
results for the two-particle species are very similar to each other.  They are particularly
inconclusive for $\sigma_m^{\textrm{ME}}$, which is seen to increase with increasing system size for
all systems except the two largest ones, for which it is seen to decrease. On the other hand,
$\sigma_m^{\textrm{DE}}$ and $\sigma_m^{\textrm{GME}}$ decrease for all system sizes, though the
results for the largest two system sizes suggest that they may saturate. This leaves open the
question of whether $\sigma_m^{\textrm{stat}}$ vanishes in the thermodynamic limit or whether it
remains finite. What is clear from the fact that both the grand-canonical ensemble and the GGE fail
to describe $m^b_k$ after relaxation \cite{Gramsch2012} is that neither eigenstate thermalization nor
generalized eigenstate thermalization take place in the HCB system in this regime.

\subsection{Quenches to the critical point}
For completeness, we present here results for the ensemble expectation values of observables in
quenches to the critical point. We note that in light of the results of Sec.~\ref{Sec:fermions} and
Ref.~\cite{Gramsch2012}, we might expect the critical regime to be particularly sensitive to
finite-size effects.

\begin{figure}[!b]
\centering
\includegraphics[width=0.475\textwidth, clip]{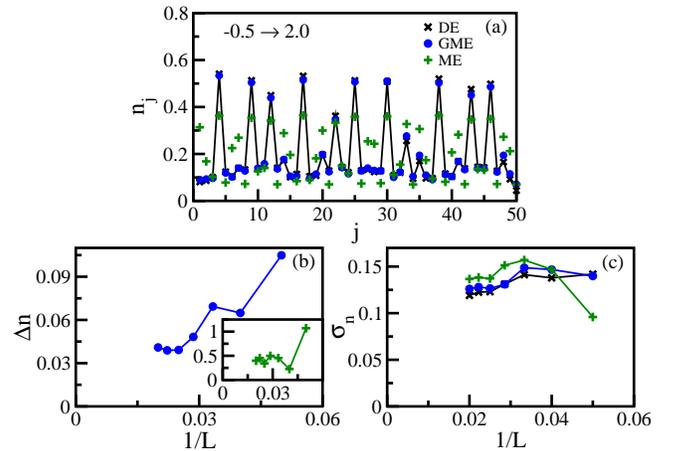}
\caption{(Color online) (a) Density profiles ($L=50$), (b) $\Delta n^{\textrm{stat-DE}}$, and
(c) $\sigma_n^{\textrm{stat}}$ in quenches to the critical point, $\lambda_I =-0.5\rightarrow\lambda_F=2.0$.
In (b), the main panel depicts $\Delta n^{\textrm{GME-DE}}$, while the inset depicts
$\Delta n^{\textrm{ME-DE}}$. The legends for the ensembles considered are the same in all panels and
are reported in (a). The values of $\delta_\textrm{GME}$ and $\delta_\textrm{ME}$ are the
same for all $L$'s and are equal to the ones in Fig.~\ref{fig:consquant}.}
\label{fig:denCri}
\end{figure}

In Fig.~\ref{fig:denCri} we present results for the density profiles, which are intermediate between those observed for quenches
to the delocalized and localized phases, as expected due to the small finite sizes of the lattice systems
studied. In particular, the average variance $\sigma_n^{\textrm{stat}}$ decreases with increasing
system size but has a tendency to saturate, so much larger system sizes will be needed to resolve whether
it vanishes in the thermodynamic limit (as it may in quenches to the delocalized phase) or whether it
remains finite (as expected from the observed behavior in quenches to the localized phase).

\begin{figure}[!t]
\centering
\includegraphics[width=0.475\textwidth, clip]{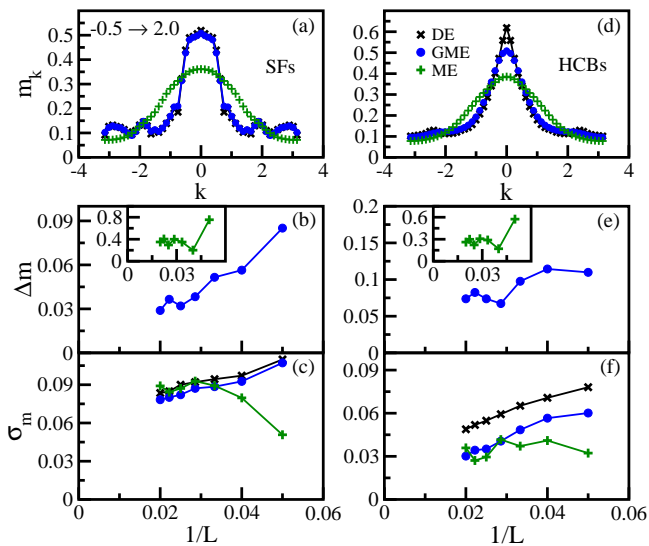}
\caption{(Color online)  [(a),(d)] Momentum distribution functions ($L=50$),
[(b),(e)] $\Delta m^{\textrm{stat-DE}}$, and [(c),(f)] $\sigma_m^{\textrm{stat}}$, for SFs (left column)
and HCBs (right column), in quenches to the critical point, $\lambda_I =-0.5\rightarrow\lambda_F=2.0$.
In [(b),(e)], the main panels depict $\Delta m^{\textrm{GME-DE}}$, while the insets depict
$\Delta m^{\textrm{ME-DE}}$. The legends for the ensembles considered are the same in all panels and
are reported in (d). The values of $\delta_\textrm{GME}$ and $\delta_\textrm{ME}$ are the
same for all $L$'s and are equal to the ones in Fig.~\ref{fig:consquant}.}
\label{fig:momCri}
\end{figure}

Figure~\ref{fig:momCri} shows results for the momentum distribution functions of SFs (left columns) and
HCBs (right columns). They are also intermediate between those obtained in the delocalized and localized
regimes. The average variance $\sigma_m^{\textrm{stat}}$ is larger for SFs than for HCBs, as in the delocalized
phase. It can also be seen to decrease with increasing
system size (as it does in the localized phase), but has a tendency to saturate (as it does in the
delocalized phase), so results for much larger system sizes will be needed to resolve whether it vanishes or remains
finite in the thermodynamic limit.

\section{summary}\label{Sec:summary}

We have studied the dynamics of the momentum distribution function of noninteracting spinless
fermions following quenches to the delocalized, localized, and critical regimes of a quasiperiodic
lattice system. We found that although the time-averaged value of this observable agrees exactly with
the predictions of the GGE in all three regimes, it does not exhibit relaxation after a quench in the
delocalized phase. This is complementary to the failure of the on-site density of hard-core bosons
(and therefore the noninteracting fermion system considered here) to equilibrate in the localized
regime that was previously observed in Ref.~\cite{Gramsch2012}. These behaviors can be understood in
terms of localization of the single-particle eigenstates of the fermion model, in momentum space for
the delocalized regime, and in real space for the localized regime, as discussed in
Refs.~\cite{cazalilla_iucci_12,ziraldo_silva_12,ziraldo_santoro_13}. This analysis also helps us
understand the previously observed failure of the GGE to describe the momentum distribution functions
of HCBs after relaxation in the localized regime \cite{Gramsch2012} as a consequence of nonvanishing
time fluctuations of one-particle fermionic correlations that persist in the thermodynamic limit.

We found that in the delocalized and localized regimes, the SF observables that do exhibit relaxation
to the GGE---the density in the former case,  and the momentum distribution function in the
latter---do so in a manner consistent with the Gaussian equilibration picture of Campos~Venuti and
Zanardi~\cite{campos_zanardi_13}. In quenches to the critical point of the Aubry-Andr\'e model, we
observed that both the density and the momentum distribution function of SFs exhibit equilibration to
the GGE, but that the decay of the time fluctuations of these quantities with increasing system size
is slower than that predicted by the conjecture of Ref.~\cite{campos_zanardi_13}.

We also studied the expectation values of one-body observables in the many-body eigenstates of the SF and HCB
Hamiltonians, comparing results for the diagonal, microcanonical, and generalized microcanonical ensembles.
We found a clear distinction between the predictions of the ME for the expectation values and those of the GME.
The differences between the expectation values in the former ensemble and the DE were consistently larger (and were
in fact greater than 10\% in all cases except for the momentum distributions in quenches to the localized phase)
and indications were found that these differences approach nonzero values as the system size is increased toward
the thermodynamic limit.  The predictions of the GME were found to be much closer to those of the DE and, for
the system sizes studied, the differences between the expectation values in these two ensembles were
observed in most cases to decrease with increasing system size.

Our study indicates that the single-particle localization---in momentum space in the delocalized
regime and in real space in the localized regime---that precludes relaxation of the corresponding
observables to the GGE also leads to finite thermodynamic-limit variances of momentum and site
occupations, respectively, in the many-body eigenstates of the SF Hamiltonian. By contrast, the
failure of the GGE to describe the momentum distribution of HCBs after relaxation in the localized
phase was not found to be associated with any corresponding saturation of the variance of momentum
occupations in eigenstates of the HCB Hamiltonian with increasing system size.

Because of finite-size effects, we did not find clear indications of the behavior of the maximum
differences between expectation values of the density and momentum distributions in distinct
eigenstates contributing to the various ensembles as the system size is increased. It would be
particularly important to understand whether the values of observables in the individual eigenstates
constituting the GME approach their average values in this ensemble with increasing system size, in
order to clarify whether generalized eigenstate thermalization occurs in the delocalized
regime---which would explain why the GGE works there for describing the momentum distribution
functions of HCBs after relaxation---and whether generalized eigenstate thermalization fails (as we
expect) in the localized regime, where the GGE fails.  It would also be interesting to see how the
addition of nearest neighbor interactions~\cite{tezuka_garcia_10, tezuka_garcia_12,
iyer_oganesyan_13}, which break integrability, modify our findings for both SFs and HCBs.

\begin{acknowledgments}
This work was supported by the Office of Naval Research (K.H., L.F.S., and M.R.), by NSF Grant
No.~DMR-1147430 (L.F.S.), by ARC Discovery Project Grant No.~DP110101047 (T.M.W.), and partially
under KITP NSF Grant No.~PHY11-25915 (L.F.S., T.M.W., and M.R.).
\end{acknowledgments}


\begin{thebibliography}{59}
\bibitem[{not()}]{noteMeanabsolute} \bibinfo{note}{The time averages of the normalized absolute
    differences
  calculated here are a lower bound to the square roots of the normalized
  variances considered in Ref.~\cite{campos_zanardi_13}.}
\end{thebibliography}
\end{document}